\documentclass[11pt,a4paper]{article}
\usepackage{graphicx}
\usepackage{epstopdf, epsfig}
\usepackage{natbib}
\usepackage{color}
\usepackage[dvipsnames]{xcolor}
\usepackage{amssymb}
\usepackage{amsmath}
\usepackage{bm} 
\usepackage{latexsym}

\newcommand\Reytau{\mathrm{Re}_{\tau}}
\newcommand\dd{\mathrm{d}}

\title{Complete matched asymptotic expansions for velocity statistics in turbulent channels}

\author{Peter A. Monkewitz\\
Ecole Polytechnique F\'ed\'erale de Lausanne (EPFL)\\ CH-1015, Lausanne, Switzerland\\
email: peter.monkewitz@epfl.ch}

\begin{document}

\maketitle   

The first complete high fidelity matched asymptotic expansions (abbreviated MAE's) are developed for all the first and second order turbulent velocity statistics in channel flow from 11 direct numerical simulations (abbreviated DNS). To put the crucial identification of overlaps on a solid footing,
a simple a priori test is devised, which only requires a DNS or experimental profile and the presumed overlap of the MAE for the quantity in question.

This test fully supports the form $c_0 - c_1\,Y^{1/4}$ of the overlaps for the stream-wise and cross-stream normal stresses $\langle uu\rangle$ and $\langle ww\rangle$, which has been advocated by \citet{chen_sreeni2022,chen_sreeni2023,chen_sreeni2025} and \citet{monkewitz22,Monkarxiv23}, recently supported by \citet{arosemena2026}, and  designated ``CS'' throughout the paper.
The first MAE analysis of the wall-normal stress $\langle vv\rangle$ then reveals a $\Reytau^{-5/4}$ scaling, i.e. an overlap of the form $c_0 - c_1\,Y^{5/4}$, which is extensively documented.
These overlaps and their domains of validity are expected to lead to improved structural models of wall turbulence.

Finally the logarithmic indicator function $\Xi \equiv y\,\dd U/\dd y$ for the mean velocity profile (abbreviated MVP) is reanalyzed, with special attention devoted to its oscillatory approach to the logarithmic MVP overlap. The latter is compared to the spatial oscillations of $\langle uu\rangle$ in the concluding section, together with further observations and suggestions.

\section{\label{sec1}Introduction}

Exploring the large Reynolds number behavior of mean and fluctuating velocities in turbulent wall-bounded flows has been an area of active and often controversial research since the seminal works of \citet{Prandtl25} and \citet{vonKarman30,vonKarman31} \citep[see for instance][for a review of early developments]{FF96}. Jumping ahead, the research aimed at understanding wall turbulence has been significantly advanced in the last 20 years by the availability of high quality DNS data. However, there is still no general agreement on the Reynolds number scaling of the most basic turbulence statistics, such as turbulent stresses.

The present paper focusses on the variance of the fluctuating velocities, but includes also another look at the mean velocity. In order to avoid the thorny separation of Reynolds number effects from geometry effects, the study concentrates on channel flow, for which the largest and most thoroughly verified DNS data base exists. Only a brief comparison to pipe flow is presented in the appendix, and zero pressure gradient turbulent boundary layers are omitted entirely, as the situation is further complicated by the presence of the turbulent-non-turbulent interface \citep[see e.g.][]{Chauhan14a}.

In the following, all velocities are inner scaled with the friction velocity $\widehat{u}_\tau \equiv (\widehat{\tau}_{\mathrm{wall}}/\widehat{\rho})^{1/2}$, where ``hats'' denote dimensional quantities. All lower case coordinates are inner scaled, without
``+''-superscripts to simplify notation, while upper case coordinates are outer-scaled. The Reynolds number $\Reytau$, finally, is the friction Reynolds number based on channel half-width throughout the paper.

The recent efforts to understand the Reynolds number scaling of moments in wall turbulence, in particular of the normal stresses $\langle uu\rangle$ and $\langle ww\rangle$, have been dominated by two ``schools'' or approaches:\newline

$\,\,\bullet\,\,$ \textit{The ``attached eddy'' (abbreviated AE) model} \newline
The model has been developed from the ``attached eddy hypothesis'' of \cite{Towns56} by \cite{PHC86},   \cite{PL90} and \cite{PML94}, among others, and has been reviewed by \cite{MarusicMonty19}.

\begin{figure}
\center
\includegraphics[width=0.45\textwidth]{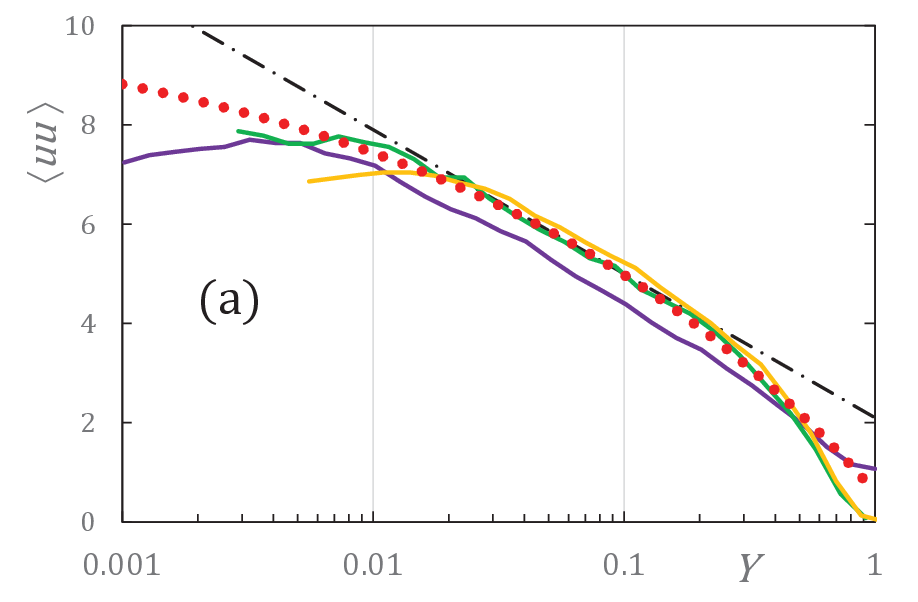}
\includegraphics[width=0.45\textwidth]{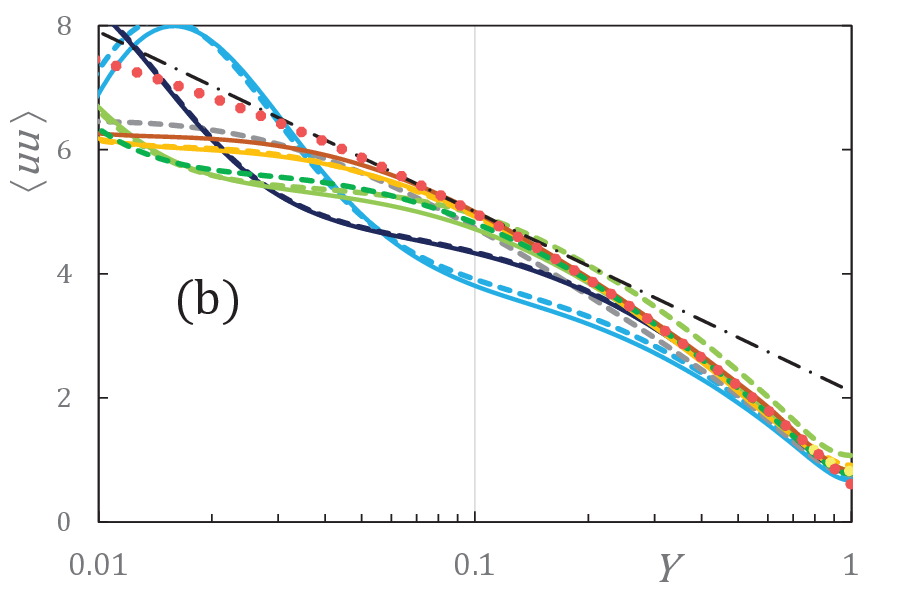}
\caption{\label{Fig1} Streamwise normal stress $\langle uu\rangle$ versus $Y$. (a) Figure 1 of \cite{MMHS13} replotted versus $Y$; --- (orange: Melbourne windtunnel; green: LCC; violet: Superpipe; SLTEST omitted).
$- \cdot - \cdot -$ (black), logarithmic fit (\ref{uuAEasymp});  $\bullet\bullet\bullet$ (red), ``bounded dissipation'' overlap (\ref{uuOLBD}).
(b) Analogue for the channel data of table \ref{TableDNS} with addition of full outer $\langle uu\rangle_{out}$ (equ. \ref{uuout} of section \ref{sec3}) (wake indicated by yellow $\bullet\bullet\bullet$ beyond $Y \approxeq 0.75$). }
\end{figure}

One of its main predictions are logarithmic asymptotes for the stream-wise and cross-stream turbulent stresses, resulting in an unbounded near-wall growth proportional to $\ln{\Reytau}$. For the stream-wise stress $\langle uu\rangle$, for instance, \cite{MMHS13} have proposed the large-$y$ behavior
    \begin{equation}
    \langle uu\rangle^{AE}(y\gg 1) = 2.1 - 1.26\,\ln{y} + 1.26\,\ln{\Reytau} \equiv 2.1 - 1.26\,\ln{Y} \quad ,
    \label{uuAEasymp}
    \end{equation}
which is indicated by black dash-dotted lines in figures \ref{Fig1}(a) and (b). The AE prediction for the  large-$y$ behavior of the wall-normal stress, on the other hand, is $\langle vv\rangle^{AE}(y\gg 1)$ = const.

To this author's knowledge, it has never been demonstrated that the large-$y$ predictions of the AE model are actually overlaps in the strict sense of MAE (see section \ref{sec2}).
Furthermore, it is already noted that both the LCC and Melbourne data are nicely fitted by the ``CS'' overlap of equation (\ref{uuOLBD}) all the way to $Y \approxeq 0.75$, well beyond $Y \approxeq 0.15$, where the data start to diverge from the logarithmic fit (\ref{uuAEasymp}). At the small-$Y$ end, the high Reynolds number experimental data in figure \ref{Fig1}(a) are not sufficiently reliable to draw any firm conclusions on where the data start to closely approach the logarithmic fit. Also, the offset of the Superpipe data will not be discussed here, as this paper concentrates on channel flow.\newline


In a series of recent papers,  \citet{chen_sreeni2021,chen_sreeni2022,chen_sreeni2023,chen_sreeni2025} have argued that the stream-wise and cross-stream normal stresses $\langle uu\rangle$ and $\langle ww\rangle$ remain bounded in the limit of $\Reytau \to \infty$, with finite Reynolds number corrections of order $\Reytau^{-1/4}$. The proposal has been motivated by the bounded dissipation of turbulence energy. Even though this rationale does not appear to be ``watertight'', the $\Reytau^{-1/4}$ scaling has received strong support from the data analyses of \cite{monkewitz22,Monkarxiv23} and the novel analysis of \cite{arosemena2026}, based on the material frame indifference of the Reynolds stress tensor.
The best fit for the ``CS'' overlap of the $\langle uu\rangle$ channel data used in this paper is
\begin{equation}
\langle uu\rangle^{CS}_{OL} = 10.6 - 10\,Y^{1/4}\quad ,
\label{uuOLBD}
\end{equation}
which is indicated by the red dotted lines in figure \ref{Fig1}. As hinted above, the exact 1/4 power in equation (\ref{uuOLBD}) cannot be rigorously justified, and more general exponents have actually been investigated by \citet{Pirozzoli_2024}, for instance. However, it is difficult to imagine how all the other equations containing normal stresses could be balanced with an ``odd'' exponent of $Y$ in equation (\ref{uuOLBD}). Besides, limiting the exponents of $Y$ to multiples of 1/4 does not appear to degrade any of the overlaps developed in this paper.
\newline

A masterful and extensive account of past research on the subject has been given in section 1 of \citet{chen_sreeni2023}, henceforth referred to as CS23, and the reader is referred to this paper for more background. While these authors have proposed and discussed the ``CS'' structure of the $\langle uu\rangle$, $\langle ww\rangle$ and $\langle pp\rangle^{1/2}$ overlaps in earlier papers, the near-perfect Reynolds number independence of these quantities between wall and respective peaks, when normalized by their respective peak values, is new. In technical MAE terms, this collapse is equivalent to the proportionality of $f_0$ and $f_1$ in equation (3.9) of CS23, but its implications for the modelling of near-wall turbulent structures remain to be explored.
Also, no complete asymptotic expansions are given in CS23 nor in other papers - So far, the only such expansion is the one by \citet{monkewitz22} for $\langle uu\rangle$.

The present paper is devoted to turbulent channels, as the largest number of DNS from different sources are available for this ``canonical'' flow.
Section \ref{sec2} starts with a brief review of MAE \citep[see e.g.][etc.]{KC85, WilcoxP}, focussed on the identification of overlaps, and proposes a simple and effective diagnostic to distinguish between overlaps and mere fits of DNS profiles. Section \ref{sec3} then presents new 2-term inner and 1-term outer expansions for $\langle uu\rangle$, which are significantly improved relative to \citet{monkewitz22} and explicitly reveal, for the first time, the scales of its inner spatial oscillations.
Then, the first complete inner, outer and composite expansions for $\langle ww\rangle$ and $\langle vv\rangle$ are developed in sections \ref{sec4} and \ref{sec5}.

In section \ref{sec6}, a new MAE description of the log indicator function $\Xi\equiv y\,(\dd U/\dd y)$ for the mean velocity $U$ is developed. The motivation for including the MVP $\Xi$ in this paper is the first characterization of its slope oscillations, which are compared to those of $\langle uu\rangle$ in the concluding section \ref{sec7}. This last section also regroups a summary of the overlap locations for the three normal stresses,  a proposal for testing the overlap scaling of $\langle vv\rangle$ and speculations on the origin of the strong flow dependence of the slope of the outer quasi-linear parts of channel, pipe and ZPG TBL MVP's.

The pressure, on the other hand, will not be discussed in this paper. Even though its overlap and near-wall collapse are rather convincing in CS23, the implications of the pressure decomposition into ``rapid'' and ``slow'' components, related respectively to squares of perturbation velocities, and a ``Stokes'' component, related linearly to velocity \citep{Pantonetal2017}, are not considered to be sufficiently understood. \newline

For all the following analyses and the construction of complete matched inner-outer asymptotic expansions, the 11 channel DNS, listed in table \ref{TableDNS}, are used.

Finally, the stream-wise normal stress in pipes is briefly presented in the appendix section \ref{app} for comparison with the channel.

\begin{table}
\center
\caption{Channel DNS profiles considered in the present study, with the line style/color scheme used in all figures }
\begin{tabular}{l r l l l}
\hline
No. & $\Reytau$ & Reference & line style/color \\
\hline
\#1& 944 & \citet{HJ06} & \fcolorbox{white}{cyan}{$\quad\quad\ \ \ $} \\
\#2& 1001 & \citet{LM15} & \fcolorbox{white}{cyan}{$\quad$}\ \fcolorbox{white}{cyan}{$\quad$} \\
\#3& 1995 & \citet{LM15} & \fcolorbox{white}{blue}{$\quad$}\ \fcolorbox{white}{blue}{$\quad$} \\
\#4& 2003 & \citet{HJ06} & \fcolorbox{white}{blue}{$\quad\quad\ \ \ $} \\
\#5& 3996 & \citet{Kaneda_Yamamoto_2021} & \fcolorbox{white}{lime}{$\quad\quad\ \ \ $} \\
\#6& 4179 & \citet{LJ14} & \fcolorbox{white}{lime}{$\quad$}\ \fcolorbox{white}{lime}{$\quad$} \\
\#7 & 5186 & \citet{LM15} & \fcolorbox{white}{green}{$\quad$}\ \fcolorbox{white}{green}{$\quad$} \\
\#8 & 7987 & \citet{Kaneda_Yamamoto_2021} & \fcolorbox{white}{yellow}{$\quad\quad\ \ \ $} \\
\#9 & 8016 & \citet{Yamamoto2018} & \fcolorbox{white}{yellow}{$\quad$}\ \fcolorbox{white}{yellow}{$\quad$} \\
\#10 & 10049 & \citet{HoyasOberlack2022} & \fcolorbox{white}{brown}{$\quad\quad\ \ \ $} \\
\#11 & 15994 & priv. comm. Y. Yamamoto& \fcolorbox{white}{gray}{$\quad$}\ \fcolorbox{white}{gray}{$\quad$} \\
\label{TableDNS}
\end{tabular}
\end{table}

\section{\label{sec2} A brief review of MAE and a simple test to identify overlaps in DNS profiles}

In order to fix ideas and nomenclature, the essentials of MAE \citep[see e.g.][etc.]{KC85,WilcoxP} are briefly reviewed with a presentation that is tailored to the problem at hand.

Here, only the basic case of two scaling regions is considered: an inner region, where the quantity of interest is described by a function $f_{in}(y)$ varying on the ''fast`` inner scale $y$, and an outer region, where it is described by $f_{out}(Y)$, which varies on the ''slow`` outer scale $Y = \epsilon y$. Here and in the following, $f_{in}$ and $f_{out}$ represent asymptotic expansions in terms of a small parameter $\mu(\epsilon)$.

To provide a description over the entire interval of interest, $f_{in}$ and $f_{out}$ need to be asymptotically matched across an overlap.
In simple cases, such as the ones being considered here, this is achieved by requiring
\begin{equation}
f_{in}(y \to \infty) = f_{out}(Y \to 0) \equiv f_{OL}   \label{match1}
\end{equation}
In more delicate cases, however, it is necessary to introduce an intermediate variable $\eta$ for the matching, with $y \gg \eta \gg Y$.

From the three $f$'s in (\ref{match1}), a composite function
\begin{equation}
f_{comp}(y) \equiv f_{in}(y) + f_{out}(Y) - f_{OL} \approxeq f_{DNS}   \label{match2}
\end{equation}
can be constructed to describe $f_{DNS}$, with the fidelity of the description depending on the number of terms carried in the expansions $f_{in}$ and $f_{out}$. For the quantities analyzed in this paper, one or two terms prove sufficient to obtain asymptotic representations within the numerical uncertainty of $f_{DNS}$.

The decomposition
\begin{equation}
f_{out}(Y) = f_{OL}(Y) + W(Y) \quad ,  \label{match3}
\end{equation}
into overlap and wake, together with relations (\ref{match1}-\ref{match2}), finally leads to
\begin{equation}
[f_{DNS} - f_{OL}](y) \cong [f_{in} - f_{OL}](y) + W(Y)  \label{match4}
\end{equation}

\subsection{\label{sec21}A simple test to identify overlaps}
The relation (\ref{match4}) provides a stringent and simple means of testing whether an assumed overlap $f_{OL}$ is indeed an overlap between inner and outer asymptotic expansions of some quantity $f$, where $f_{OL}$ includes all higher order terms necessary for a satisfactory characterization of the overlap.

\textit{Provided} $f_{OL}$ is the correct overlap, the difference $[f_{in} - f_{OL}](y)$, by definition (\ref{match1}) of overlaps, goes smoothly to zero at the inner start of the overlap $y_{OL\,start}$. For $y > y_{OL\,start}$, the difference $[f_{in} - f_{OL}](y)$ remains ``zero'', i.e. smaller than the uncertainty of $f_{DNS} - f_{OL}$, until $W(Y)$ becomes appreciable beyond some $Y_{W\,start}$. Hence, the only condition for this simple test to ``work'' is that $y_{OL\,start} < (\Reytau\,Y_{W\,start})$, i.e. that $\Reytau$ is sufficiently large.

There are two additional essential conditions to qualify $f_{OL}$ as an overlap:
\begin{itemize}
\item $\vert f_{DNS} - f_{OL}\vert$ must, for all $\Reytau$, drop below its uncertainty level at a $y_{OL\,start}$ which is independent of $\Reytau$,
\item and $\vert f_{DNS} - f_{OL}\vert$ must remain below this level up to a fixed $Y_{W\,start}$, beyond which the wake in equation (\ref{match3}) becomes appreciable. \newline
\end{itemize}

\begin{figure}
\center
\includegraphics[width=0.6\textwidth]{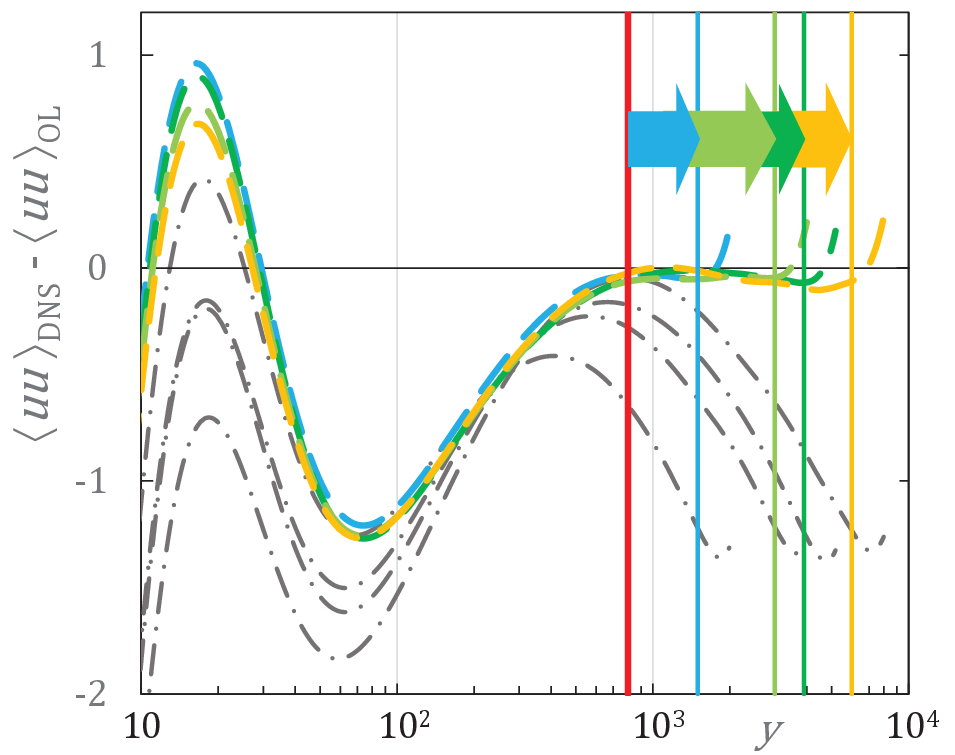}
\caption{\label{Fig2} Difference between the $\langle uu\rangle_{DNS}$ profiles \#3, 5, 7 and 8 of table \ref{TableDNS} and the ``CS'' overlap (\ref{uuOLBD}) (color $- - -$), as well as the logarithmic law (\ref{uuAEasymp}) (grey $- \cdot - \cdot -$). Vertical red line: start of the ``CS'' overlap at $y \approxeq 800$; vertical colored lines: approximate end of the ``CS'' overlap at $Y\approxeq 0.75$. Horizontal arrows: extent of overlap for the four $\Reytau$.}
\end{figure}

In figure \ref{Fig2} both overlaps (\ref{uuAEasymp}) and (\ref{uuOLBD}) for $\langle uu\rangle$ are tested. To avoid clutter, only four profiles of table \ref{TableDNS} are shown and all clearly demonstrate that the logarithmic fit (\ref{uuAEasymp}) does not pass the overlap test, while the overlap (\ref{uuOLBD}) clearly passes.
The same conclusion is reached with modified coefficients in (\ref{uuAEasymp}) (not shown in fig. \ref{Fig2}).
The test (\ref{match4}) will be demonstrated to be also effective to validate the overlaps of $\langle ww\rangle$ and $\langle vv\rangle$ in sections \ref{sec4} and \ref{sec5}.

When dealing with experimental profiles, the uncertainties are typically at least one order of magnitude larger than in DNS, making the identification of overlaps much more challenging. In a recent paper, \citet{Nagib_Marusic_2025} have analyzed a large number of data by dividing them by various logarithmic and power-law fitting relations.
They have however not made clear which ones are proposed as overlaps. Their choice of  starting points $y_{in} \sim \Reytau^{1/2}$ for the ZPG and Superpipe fits actually disqualifies them as overlaps, which \textit{must} start at a fixed value of the inner coordinate $y$, independent of $\Reytau$.

For perfect data, validating an overlap by subtracting it from the data and by forming the ratio of data and presumed overlap are trivially equivalent.
In practice, however, the $y$-dependence of the data uncertainty must be considered:
\begin{itemize}
\item In DNS, the step size generally increases with distance from the wall, while gradients decrease, leaving the uncertainty very roughly independent of wall distance. In this situation, subtracting the overlap from the data as in equation (\ref{match4}) is preferable, as the overlaps considered here are decreasing functions of $y$ and a division would amplify the uncertainty towards the centerline.
\item In experimental data, on the other hand, the uncertainty often increases dramatically towards the wall, and dividing the data by the presumed overlap may provide a more balanced view of the difference between data and overlap.
\end{itemize}

Another effective method to test data for the presence of a particular scaling range is the indicator function. A logarithmic region of the form (\ref{uuAEasymp}), for instance, is revealed by the indicator function $\Xi_{\mathrm{log}}= Y \dd \langle uu\rangle/\dd Y$, which takes on a constant value of -1.26 in the interval where the data obey equation (\ref{uuAEasymp}). A region described by the power law (\ref{uuOLBD}), on the other hand, corresponds to a constant value of the indicator function $\Xi_{\mathrm{BD}}= 4\,Y^{3/4} \dd \langle uu\rangle/\dd Y$, equal to $-10$. However, a region of constant indicator function is \textit{not sufficient} to identify overlaps, as there are strict rules for the Reynolds number scaling of their start- and end-points (see above).

\begin{figure}
\center
\includegraphics[width=0.48\textwidth]{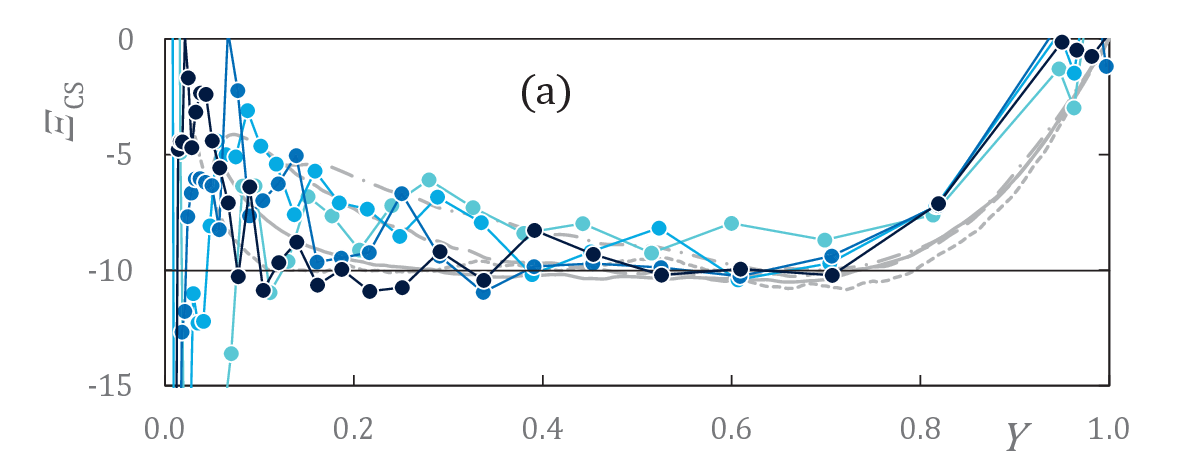}
\includegraphics[width=0.48\textwidth]{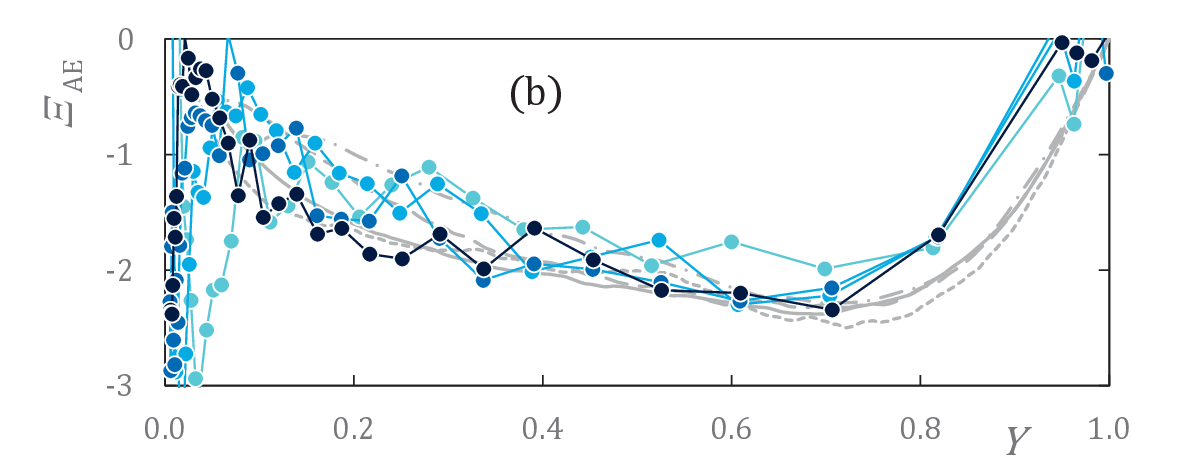}
\caption{\label{figXich} Indicator functions $\Xi_{\mathrm{CS}}= 4\,Y^{3/4} \dd \langle uu\rangle/\dd Y$  in panel (a) and $\Xi_{\mathrm{AE}}= Y \dd \langle uu\rangle/\dd Y$ in panel (b) for the experimental channel data of \citet{SchultzFlack2013}, with $\Reytau = 1010, 1960, 4040, 5900$ (increasingly dark blue $-\bullet-$). Grey lines: reproduction, for comparison, of the four channel indicator functions for profiles \#2, 3 7 and 10 of table \ref{TableDNS}.}
\end{figure}

As an example, these two indicator functions are shown in figure \ref{figXich} for the laser Doppler channel data of \citet{SchultzFlack2013}. While there is considerable scatter due to the differentiation of experimental data, there can be no doubt that the data closely follow the bounded dissipation scaling, i.e. show, within experimental uncertainty, a region of constant $\Xi_{CS}\approxeq -10$ in figure \ref{figXich}(a), which expands towards the wall with increasing Reynolds number. Note in particular the close correspondence between the experiment for $\Reytau=5900$ and the DNS \#7 for $\Reytau=5186$. The test for a logarithmic region in figure \ref{figXich}(b), on the other hand, shows no approach to a constant.
An analogous figure \ref{figpipe} of the two pipe indicator functions is shown and briefly discussed in the appendix section \ref{app}.

\section{\label{sec3}The stream-wise normal stress $\langle uu\rangle$}

Like the MVP, the stream-wise normal stress has been extensively discussed in the literature and the question of whether $\langle uu\rangle$ is unbounded or not for $\Reytau \to \infty$ is intimately linked to the functional form of the overlap, as discussed in the introductory section \ref{sec1}. As argued in section \ref{sec2} and figure \ref{Fig2}, the overlap of $\langle uu\rangle$ for the channel data in table \ref{TableDNS} is the ``CS'' overlap of equation (\ref{uuOLBD}), shown by the red dotted lines in figure \ref{Fig1}.

The complete outer expansion is obtained by adding a wake function $W(Y)$ to the overlap (\ref{uuOLBD}). In the following, all wakes will be modelled by the generic function $W(Y) = \Delta_1\,\exp{[-c_1\,(1-Y)]} + \Delta_2\,\exp{[-c_2\,(1-Y)^2]}$.

Hence, the complete outer expansion of $\langle uu\rangle$, shown in figure \ref{Fig1}(b),  is
\begin{eqnarray}
\langle uu\rangle_{out}(Y) &=& 10.6 - 10\,Y^{1/4} + W_{uu} \quad \mathrm{with} \label{uuout}\\
W_{uu}(Y) &=& 0.29\,\exp{[-8.621\,(1-Y)]} - 0.07\,\exp{[-10\,(1-Y)^2]} \, , \label{Wuu}
\end{eqnarray}
where the product $0.29 \times 8.621 = 10/4$ ensures a zero centerline slope of $\langle uu\rangle_{out}$.

Based on the evidence of figure \ref{Fig2}, the inner expansion, which blends into this overlap at $y \approxeq 800$,
is of the form $\langle uu\rangle = f_0(y) + \Reytau^{-1/4}\,f_1(y) \,+$ ... , with $f_0 \to 10.6$ and $f_1 \to -10\, y^{1/4}$ for $y \gg 1$. A first fit of $\langle uu\rangle$ has been given in \citet[][fig. 3]{monkewitz22}, but had several flaws: it fitted the inner maximum with an added Gaussian and used a ``patch'' at $y = 470$.

\begin{figure}
\center
\includegraphics[width=0.6\textwidth]{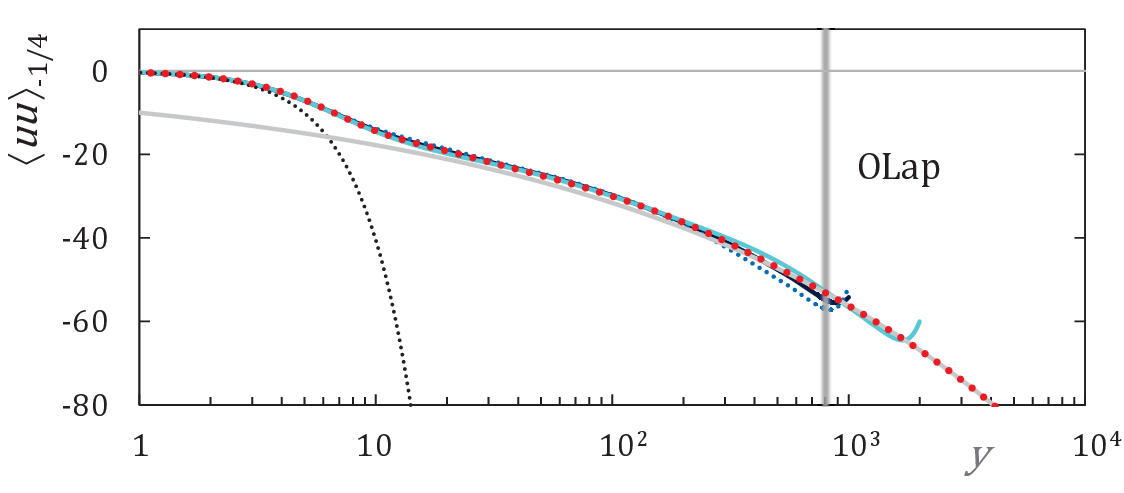}
\caption{\label{Fig3} Order $\mathcal{O}(\Reytau^{-1/4})$ of stream-wise normal stress $\langle uu\rangle$ extracted from DNS in fig. 1b of \cite{monkewitz22} \citep[see also][fig. 2]{Monkarxiv23}. $\cdot \cdot \cdot$ (black), 1st term $-0.405\,y^2$ of Taylor series; --- (grey), outer overlap $-10\,y^{1/4}$; $\bullet\bullet\bullet$ (red), new complete fit (\ref{uu025}) of the $\mathcal{O}(\Reytau^{-1/4})$ contribution to $\langle uu\rangle$. }
\end{figure}

Here, a new significantly improved, complete composite expansion for $\langle uu\rangle$ is developed in terms of exponentials,
which naturally capture the extrema of $\langle uu\rangle_{in}$ and explicitly reveal the spatial scales associated with each extremum.
In a first step, the $\mathcal{O}(\Reytau^{-1/4})$ contribution to $\langle uu\rangle$ of \citet[][fig. 1b]{monkewitz22} is refitted by
\begin{eqnarray}
\label{uu025}
\langle uu\rangle_{in,\,-1/4} = -y^{1/4}\,\{10 - 0.6\,\exp(-0.00005\,y^{7/4}) \nonumber \\
\quad\quad\quad\quad -1.0\,\exp(-0.006\,y^{7/4}) - 8.4\,\exp(-0.047\,y^{7/4})\,\}
\end{eqnarray}
and shown in figure \ref{Fig3}. The above coefficients yield the best fit Taylor expansion  $\langle uu\rangle_{in,\,-1/4} = -0.405\,y^2 + ...\,$.

\begin{figure}
\center
\includegraphics[width=0.6\textwidth]{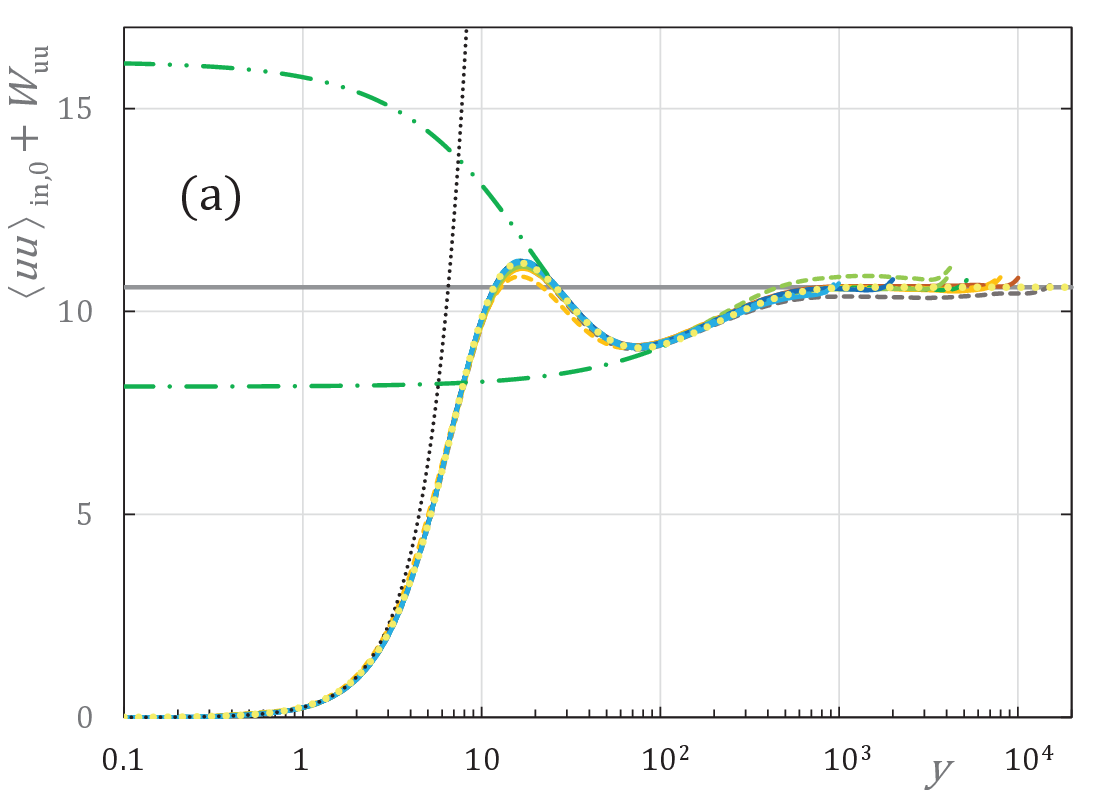}
\includegraphics[width=0.286\textwidth]{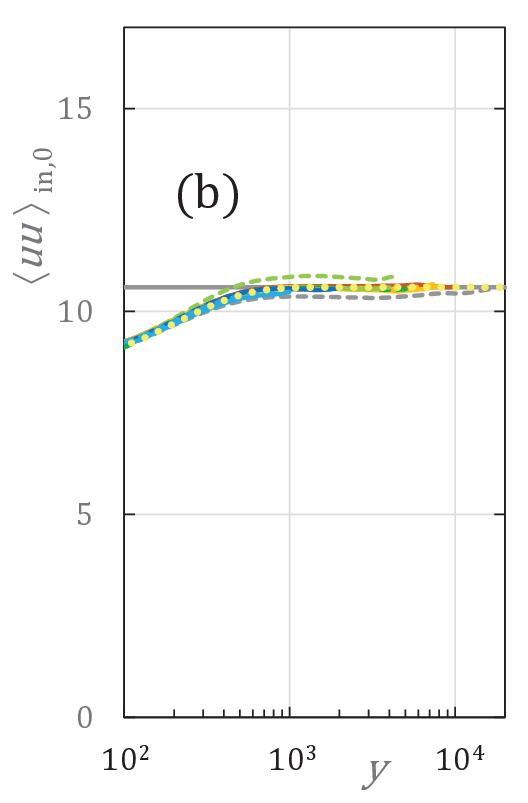}
\caption{\label{Fig4} Order $\mathcal{O}(1)$ of stream-wise normal stress $\langle uu\rangle_{in, 0}$ (\ref{uu01}-\ref{uu03}) plus wake (\ref{Wuu}) in panel (a) and without wake in (b). $\cdot \cdot \cdot$ (black), 1st term $(1/4)\,y^2$ of Taylor series; --- (grey), small-$Y$ limit $10.6$ of overlap; $- \cdot - \cdot -$ (green), exponential (\ref{uu01}); $- \cdot \cdot - \cdot \cdot -$ (green), exponential (\ref{uu02}); $\bullet \bullet \bullet$ (light yellow), complete inner fit (\ref{uu01}) - (\ref{uu03}) of the $\mathcal{O}(1)$ contribution to $\langle uu\rangle$. }
\end{figure}

Subtracting the order $\mathcal{O}(\Reytau^{-1/4})$ of equation (\ref{uu025}) from the DNS yields the leading order of the inner expansion $\langle uu\rangle_{in, 0}$, shown in figure \ref{Fig4}. The large majority of data are seen to collapse very nicely onto the sum of exponentials
\begin{eqnarray}
\langle uu\rangle_{in, 0} &=& 10.6 - 2.45\,\exp(-0.005\,y) \label{uu01}\\
   &+&8\,\exp(-0.05\,y) \label{uu02}\\
   &-&16.15\,\exp(-0.02401\,y\,-0.01639\,y^2\,-1.0317\,10^{-5}\,y^3 \nonumber \\
   &\,& \quad\quad\quad\quad\quad  +4.1\,10^{-5}\,y^4\,-1.1\,10^{-6}\,y^5)\quad , \label{uu03}
\end{eqnarray}
where the coefficients of $y$ to $y^4$ in the last exponential (\ref{uu03}) are determined such as to produce the best fit Taylor expansion of $\langle uu\rangle_{in, 0}$ about the wall, starting with $(1/4)\,y^2$.

The new description of the damped oscillations of $\langle uu\rangle_{in, 0}$ in terms of simple exponentials has the unique advantage of providing a straightforward characterization of the spatial scales associated with each oscillation by the coefficients in the exponents: The first two exponentials (\ref{uu01}-\ref{uu02}) yield scales of $(0.005)^{-1} = 200$ and $(0.05)^{-1} = 20$ inner units. Since the last exponential (\ref{uu03}) has a complicated argument, constrained to produce the best fit Taylor expansion, the shortest scale is taken to be the intersection of the Taylor expansion with the asymptotic value 10.6 of $\langle uu\rangle_{in, 0}$ (equ. \ref{uu01}), i.e.
$(4 \times 10.6)^{1/2} \approxeq 6.5$, which is of the order of the viscous sublayer thickness. This sequence of length scales will be compared in the final section \ref{sec7} to the sequence of scales characterizing the spatial oscillations of the MVP indicator function educed in section \ref{sec6}.

\section{\label{sec4}The cross-stream normal stress $\langle ww\rangle$}

\begin{figure}
\center
\includegraphics[width=0.7\textwidth]{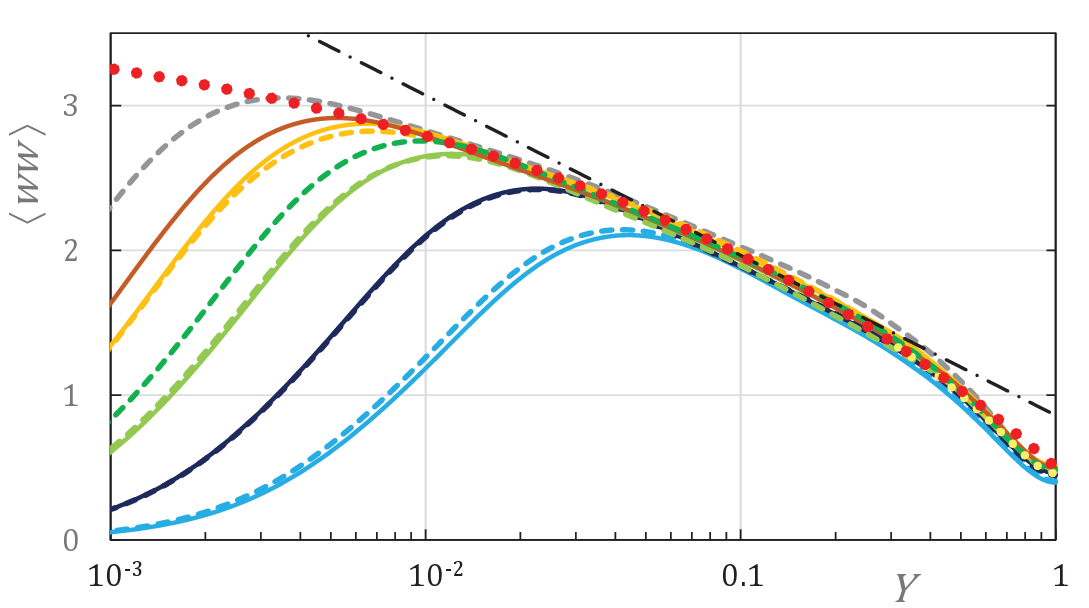}
\caption{\label{Fig5} Cross-stream normal stress $\langle ww\rangle$ versus $Y$. $\bullet \bullet \bullet$ (red), overlap (\ref{wwOLBD}); $\bullet \bullet \bullet$ (light yellow), full outer expansion (\ref{wwout}); $- \cdot - \cdot -$ (black), example logarithmic fit $0.86 - 0.48\,\ln{Y}$ . }
\end{figure}

The cross-stream stress $\langle ww\rangle$ is shown in figure \ref{Fig5} for the DNS of table \ref{TableDNS}. As for the stream-wise stress, the ``bounded dissipation'' overlap
\begin{equation}
\langle ww\rangle_{OL} = 3.85 - 3.40\,Y^{1/4}\quad ,
\label{wwOLBD}
\end{equation}
is seen in figure \ref{Fig5} to closely fit the data from $y \approxeq 60$ all the way to $Y \approxeq 0.4$, while the example of a logarithmic fit hugs the data only over the short interval $Y \in [0.1, 0.2]$.

\begin{figure}
\center
\includegraphics[width=0.6\textwidth]{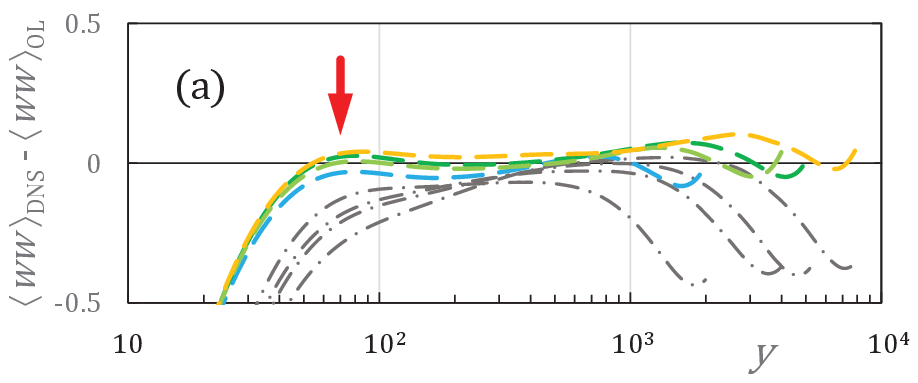}
\includegraphics[width=0.6\textwidth]{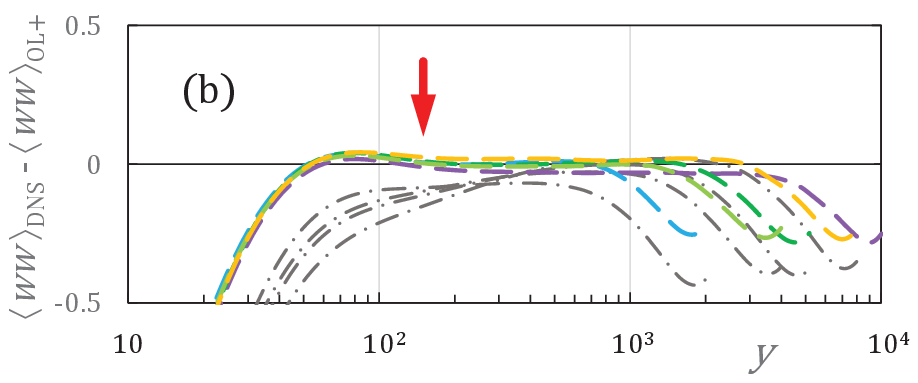}
\caption{\label{Fig6} (a) Difference between the profiles of $\langle ww\rangle_{DNS}$ \#3, 5, 7 and 8 of table \ref{TableDNS} and the ``CS'' overlap (\ref{wwOLBD}) (color $- - -$), as well as the logarithmic law $0.86 - 0.48\,\ln{Y}$ (grey $- \cdot - \cdot -$). Vertical red arrow, start of the ``bounded dissipation'' overlap at $y \approxeq 70$.
(b) Same data as in (a) plus profile \#10, minus the expanded overlap (\ref{wwOLBDplus}). Vertical red arrow, start of the overlap moved to $y \approxeq 150$}
\end{figure}

The test of overlap (\ref{wwOLBD}) devised in section \ref{sec2} is shown in figure \ref{Fig6}(a). While the ``CS'' overlap (\ref{wwOLBD}) does not look quite as convincing as the one for $\langle uu\rangle$ in figure \ref{Fig2}, it appears clearly superior to the
a logarithmic law, an example of which is included in the figure. A straightforward extension of the overlap fit (\ref{wwOLBD}) to
\begin{equation}
\langle ww\rangle_{OL+} = 3.87 - 3.40\,Y^{1/4} + 0.3\,Y - 200\,\Reytau^{-1}
\label{wwOLBDplus}
\end{equation}
actually resolves the problem almost perfectly in figure \ref{Fig6}(b), in which the \#10 profile of table \ref{TableDNS} has been added to show how the narrow computational box of this DNS affects the $\langle ww\rangle$ overlap.

Also noted in figure \ref{Fig6} is a slight ``hump'' of the ``CS'' overlap, centered at $y \approxeq 75$, and it is not clear whether it represents an actual flow feature or can be attributed to computational uncertainty. Hence, the start of the overlap can only be located between $y \approxeq 70$ and 150. This refined analysis demonstrates that equation (\ref{wwOLBD}) is a valid low-order approximation of the overlap.
To avoid developing a matching correction of order $\mathcal{O}(\Reytau^{-1})$ for the inner expansion, which is not essential to the understanding of the asymptotic structure of $\langle ww\rangle$, the basic overlap (\ref{wwOLBD}) is used in the following.
This leads to the outer expansion of $\langle ww\rangle$
\begin{eqnarray}
\langle ww\rangle_{out}(Y) = 3.85 - 3.40\,Y^{1/4} + W_{ww}(Y) \quad \mathrm{with} \label{wwout} \\
W_{ww}(Y) = 0.14\,\exp{[-6.071\,(1-Y)]} - 0.18\,\exp{[-6\,(1-Y)^2]} \, , \label{Www}
\end{eqnarray}
where the product $0.14 \times 6.071 = 3.40/4$ ensures a zero centerline slope of $\langle ww\rangle_{out}$.

As for $\langle uu\rangle$, the inner expansion of $\langle ww\rangle$ must be of the form
$\langle ww\rangle = f_0(y) + \Reytau^{-1/4}\,f_1(y)\, +$ ... , with $f_0 \to 3.85$ and $f_1 \to -3.40\, y^{1/4}$ for $y \gg 1$ in order to asymptote to the overlap (\ref{wwOLBD}). This is achieved by the simple fit for the $\Reytau^{-1/4}$ term in the inner expansion
\begin{equation}
\label{ww025}
\langle ww\rangle_{in,\,-1/4} = -3.40\, y^{1/4}\,\{1  - \exp(-0.047\,y^{7/4}) \}
\end{equation}

\begin{figure}
\center
\includegraphics[width=0.6\textwidth]{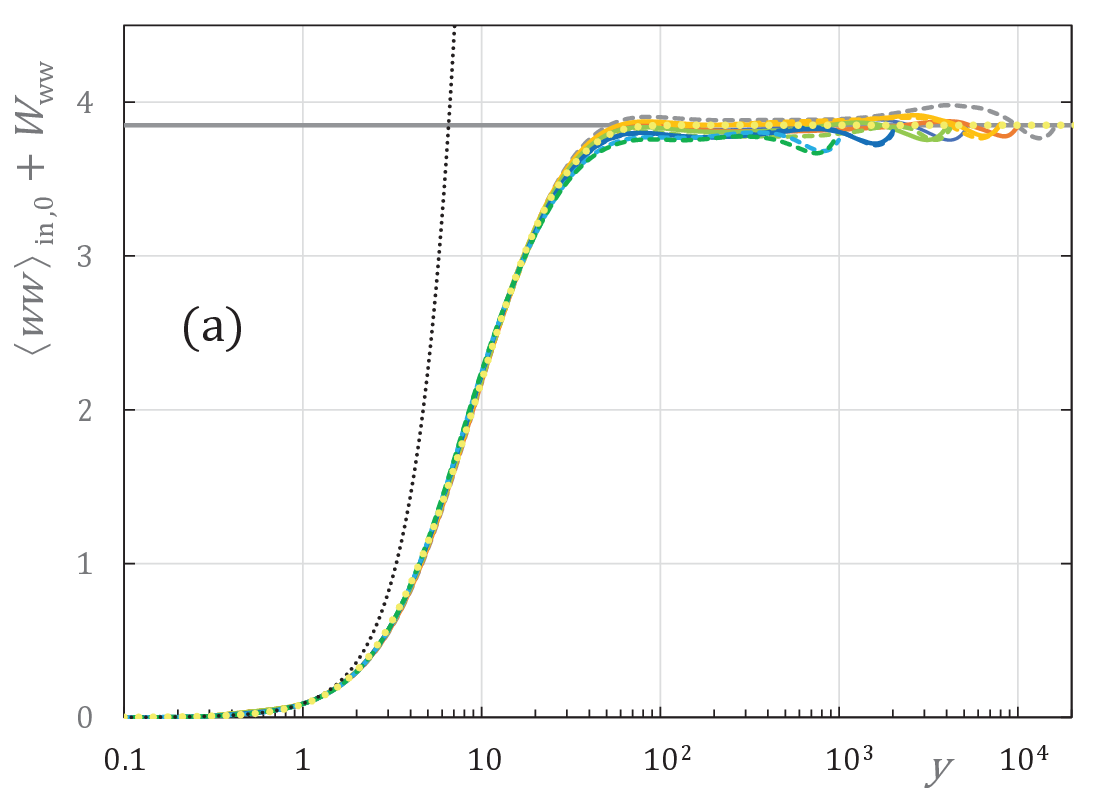}
\includegraphics[width=0.286\textwidth]{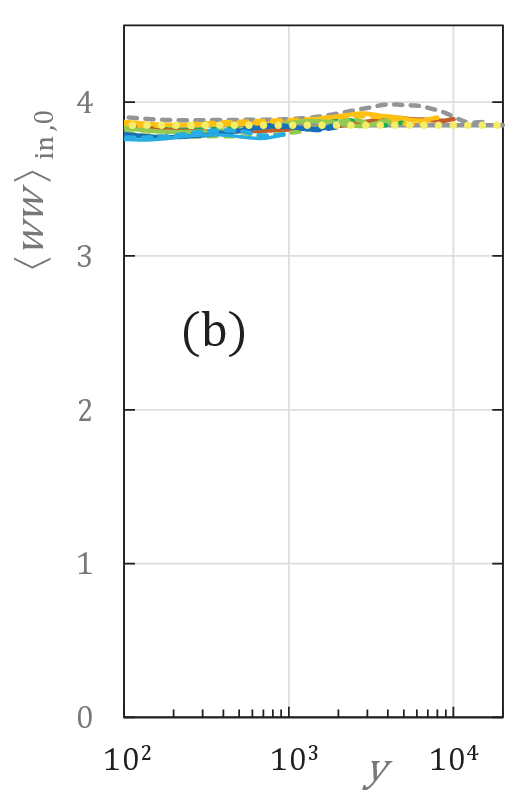}
\caption{\label{Fig7} Order $\mathcal{O}(1)$ of cross-stream normal stress $\langle ww\rangle_{in, 0}$ (\ref{ww0}) plus wake (\ref{Wuu}) in panel (a) and without wake in (b).  $\cdot \cdot \cdot$ (black), 1st term $0.09\,y^2$ of Taylor series; --- (grey), small-$Y$ limit $3.85$ of outer overlap; $\bullet \bullet \bullet$ (light yellow), complete inner fit (\ref{ww0}) of the $\mathcal{O}(1)$ contribution to $\langle ww\rangle$. }
\end{figure}

Subtracting the higher order term (\ref{ww025}) from the DNS yields the infinite Reynolds number limit $\langle ww\rangle_{in,\,0}$ of the inner expansion in figure \ref{Fig7}.

As $\langle ww\rangle_{in,\,0}$ shows no oscillations, the Pad\'e approximant (\ref{ww0}), indicated by the yellow dots in figure \ref{Fig7}, is well adapted to fit $\langle ww\rangle_{in, 0}$ :
\begin{eqnarray}
\langle ww\rangle_{in, 0} = &\{& 0.09\,y^2 + 0.0018\,y^4 + 2.\,10^{-10}\,y^8 \}\,\times \nonumber \\
   &\{& 1 + 0.067\,y^2 + 0.000465\,y^4 + (2.\,10^{-10}/3.85)\,y^8 \}^{-1}  \label{ww0}
\end{eqnarray}

With the complete outer expansion (\ref{wwout}, \ref{Www}), the full composite expansion of $\langle ww\rangle$ is finally
\begin{equation}
\label{wwcomp}
\langle ww\rangle_{comp} = \langle ww\rangle_{in, 0} + \Reytau^{-1/4}\,\langle ww\rangle_{in,\,-1/4} + W_{ww}(Y) \quad ,
\end{equation}
where $W_{ww}(Y)$ is given by equation (\ref{Www}).

\section{\label{sec5}The wall-normal stress $\langle vv\rangle$}

Compared to the stream-wise Reynolds stress, the wall-normal stress has received virtually no attention, except for the case of rough walls, where $\langle vv\rangle$ plays a key role in the interactions between ground and atmosphere \citep[see for instance][]{Orlandi2013}. There exist also very few data fits to corroborate proposed scalings - one specific proposal is the Taylor series of $\langle vv\rangle$, fitted by an expansion in $\ln{\Reytau}$ by \citet{smitsetal2021}, another the constant large-$y$ limit derived from the attached eddy model, both not supported by the following data analysis.

\begin{figure}
\center
\includegraphics[width=0.45\textwidth]{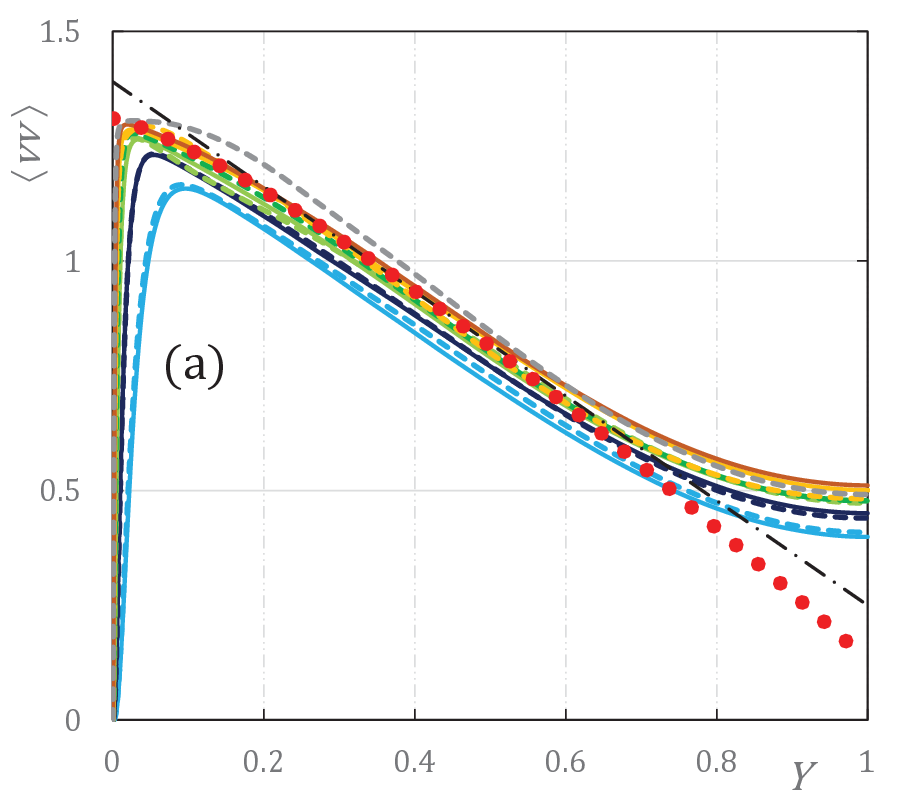}
\includegraphics[width=0.45\textwidth]{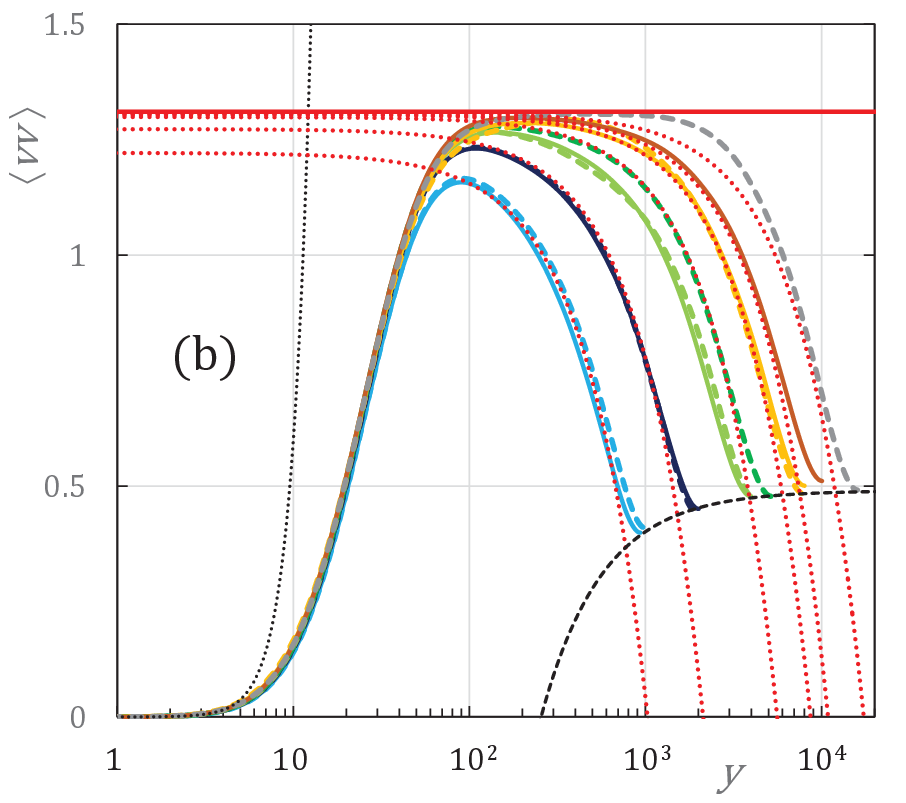}
\caption{\label{Fig8} (a) Wall-normal stress $\langle vv\rangle$ versus $Y$. $\bullet \bullet \bullet$ (red), overlap (\ref{vvOL}) for infinite $\Reytau$; $- \cdot - \cdot -$ (black), linear fit $1.39 - 1.14\,Y$ to guide the eye. \\
(b) $\langle vv\rangle$ versus $y$. $\cdot \cdot \cdot$ (black), 1st term $0.00015\,y^4$ of Taylor series; - - - (black), equ. (\ref{vvCL}) for the centerline $\langle vv\rangle_{CL}$; --- (red), overlap (\ref{vvOL}) at $\Reytau = \infty$ ; $\cdot \cdot \cdot$ (red), overlaps at selected finite $\Reytau$.}
\end{figure}

To develop the full MAE description of $\langle vv\rangle$, it is useful to first inspect the set of data in table \ref{TableDNS}, plotted against both outer and inner wall-normal coordinates in figures \ref{Fig8}(a) and (b) (note that the preliminary profile \#11 differs significantly from all the others and will in the following be disregarded for the fits).
Several features of $\langle vv\rangle$ in this figure are noteworthy: After the rise close to the wall, corresponding to the inner expansion, the decaying parts in figure \ref{Fig8}(a) are progressively shifted down as $\Reytau$ is decreased, by amounts that are to a good approximation independent of $Y$. The second significant feature of figure \ref{Fig8}(a)
is the gradual departure of $\langle vv\rangle$ from the linear fit $(1.39 - 1.14\,Y)$ below $Y \approxeq 0.15$.

This departure is reproduced by the overlap
\begin{equation}
\label{vvOL}
\langle vv\rangle_{OL} = 1.31 - 1.18\,Y^{5/4} - 500\,\Reytau^{-5/4} \quad ,
\end{equation}
which is seen in figure \ref{Fig8}(b) to nearly perfectly reproduce the $\Reytau$-dependent shifts and the negative curvature of $\langle vv\rangle$ towards the wall (again, disregarding profile \#11).

\begin{figure}
\center
\includegraphics[width=0.6\textwidth]{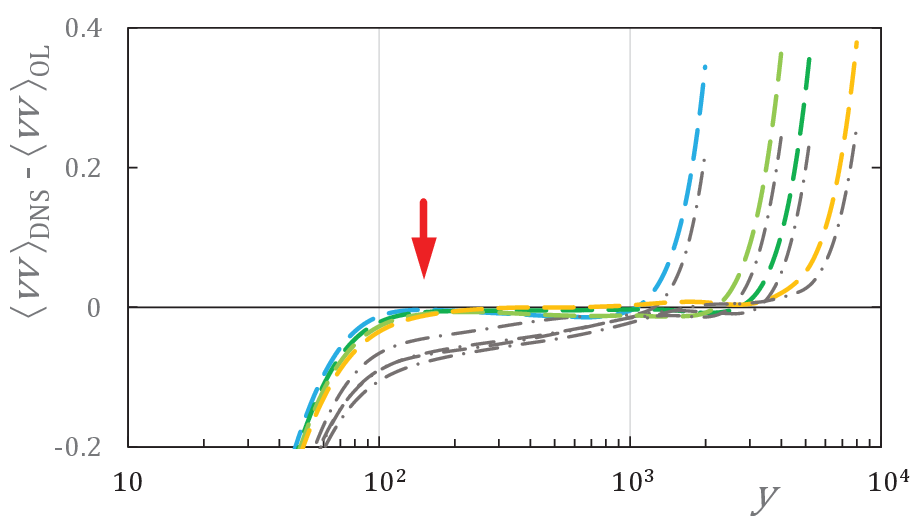}
\caption{\label{Fig9} Difference between the profiles of $\langle vv\rangle_{DNS}$ \#3, 5, 7 and 8 of table \ref{TableDNS} and the overlap (\ref{vvOL}) (color $- - -$), as well as the linear fit $1.39 - 1.14\,Y$ (grey $- \cdot - \cdot -$). Vertical red arrow, start of the overlap at $y \approxeq 150$.}
\end{figure}

The choice of the $\Reytau^{-5/4}$ scaling is again verified by the test devised in section \ref{sec2}, shown in figure \ref{Fig9}, where equation (\ref{vvOL}) is compared to the linear fit $1.39 - 1.14\,Y$ . Further support for the overlap (\ref{vvOL}) comes from the excellent fit of $\langle vv\rangle_{CL}(y=\Reytau)$ on the centerline

\begin{equation}
\label{vvCL}
\langle vv\rangle_{CL} = 0.49 - 500\,\Reytau^{-5/4} \quad ,
\end{equation}
 seen in figure \ref{Fig8}(b). Note that the difference between $1.31 - 1.18 = 0.13$ in equation (\ref{vvOL}) and 0.49 in equation (\ref{vvCL}) is the value of the wake function (\ref{Wvv}) on the centerline.
The latter is well fitted by
\begin{equation}
\label{Wvv}
W_{vv}(Y)  = 0.36\,\exp{[ -4.0972\,(1-Y) - 8\,(1-Y)^{5/2}]} \quad ,
\end{equation}
where the coefficient of $(1-Y)$ is adjusted to yield a zero centerline slope for the outer expansion $\langle vv\rangle_{out}$ , given by
\begin{eqnarray}
\langle vv\rangle_{out} &=&  \langle vv\rangle_{out,\,0} + \Reytau^{-5/4}\,\langle vv\rangle_{out,\,-5/4} \quad \mathrm{with} \label{vvout1} \\
 \langle vv\rangle_{out,\,0} &=& 1.31 - 1.18\,Y^{5/4} + W_{\langle vv\rangle}(Y)  \label{vvout2} \\
 \langle vv\rangle_{out,\,-5/4} &=& -500  \label{vvout3}
\end{eqnarray}

\begin{figure}
\center
\includegraphics[width=0.4\textwidth]{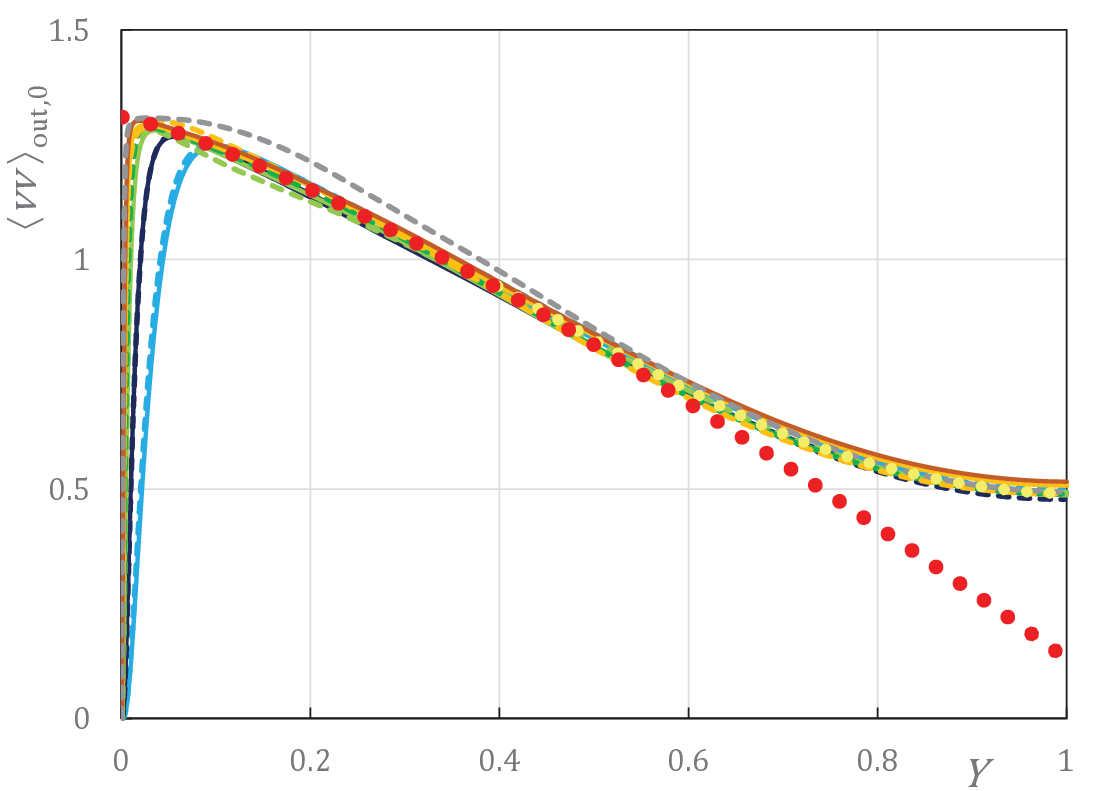}
\caption{\label{Fig10} Infinite $\Reytau$ limit of outer wall-normal stress $\langle vv\rangle_{out}$ (equ. \ref{vvout1}) versus $Y$. $\bullet \bullet \bullet$ (red), overlap (\ref{vvOL}) at infinite $\Reytau$; $\bullet \bullet \bullet$ (light yellow), infinite $\Reytau$ limit of full outer expansion (\ref{vvout2}). }
\end{figure}

Figure \ref{Fig10} shows the infinite $\Reytau$ limit $\langle vv\rangle_{out,\,0} = \langle vv\rangle_{DNS} - \Reytau^{-5/4}\,\langle vv\rangle_{out,\,-5/4}$, together with the infinite $\Reytau$ limits of the overlap \ref{vvOL} and of the full fit of $\langle vv\rangle_{out,\,0}$ given by equation (\ref{vvout2}).

Turning to the inner expansion of order $\mathcal{O}(\Reytau^{-5/4})$, the uniform outer offset (\ref{vvout3}) of order $\mathcal{O}(\Reytau^{-5/4})$ has to be brought smoothly down to zero at the wall. This is achieved with a simple Pad\'e approximant which produces the appropriate Taylor series of $\langle vv\rangle$ about the wall, starting with $y^4$. Hence, the full $\mathcal{O}(\Reytau^{-5/4})$ contribution $\langle vv\rangle_{in,\,-5/4}$ matching the overlap (\ref{vvOL}) for $y \gg 1$ is
\begin{equation}
\label{vvin1}
\langle vv\rangle_{in,\,-5/4}  = -500\,\frac{2.\,10^{-7}\,y^4}{1 + 2.\,10^{-7}\,y^4}\,- 1.18\,y^{5/4}
\end{equation}

\begin{figure}
\center
\includegraphics[width=0.6\textwidth]{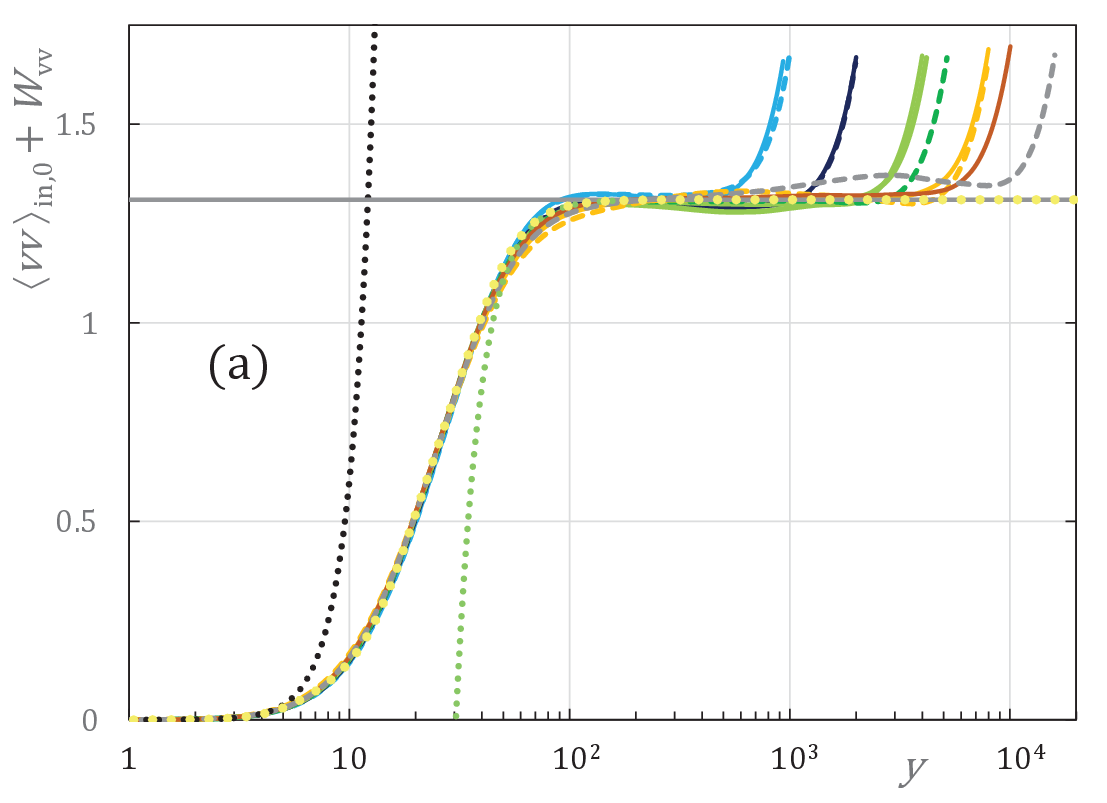}
\includegraphics[width=0.286\textwidth]{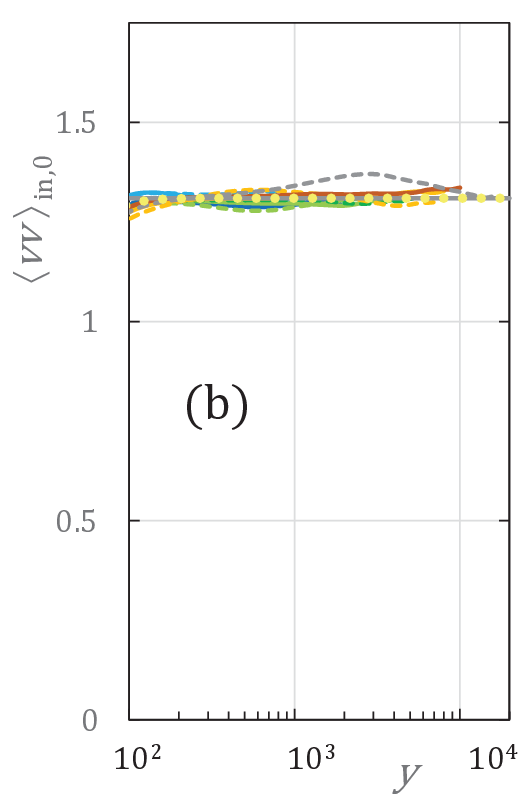}
\caption{\label{Fig11} Infinite $\Reytau$ limit of inner wall-normal stress equal to $\langle vv\rangle_{DNS} - \Reytau^{-5/4}\,\langle vv\rangle_{in,\,-5/4}$ (equation \ref{vvin1}) versus $y$ in (a), and without the wake (\ref{Wvv}) in (b). $\bullet \bullet \bullet$ (red), fit (\ref{vv0}); $\cdots$ (black), Taylor expansion $6.\,10^{-5}\,y^4$; $\cdots$ (aqua), asymptotic approach to 1.31, equal to the 1st two terms of equ. (\ref{vvmax2}), with $C = 4.81\,10^5$ and $\alpha = 15/4$ .}
\end{figure}

\begin{figure}
\center
\includegraphics[width=0.4\textwidth]{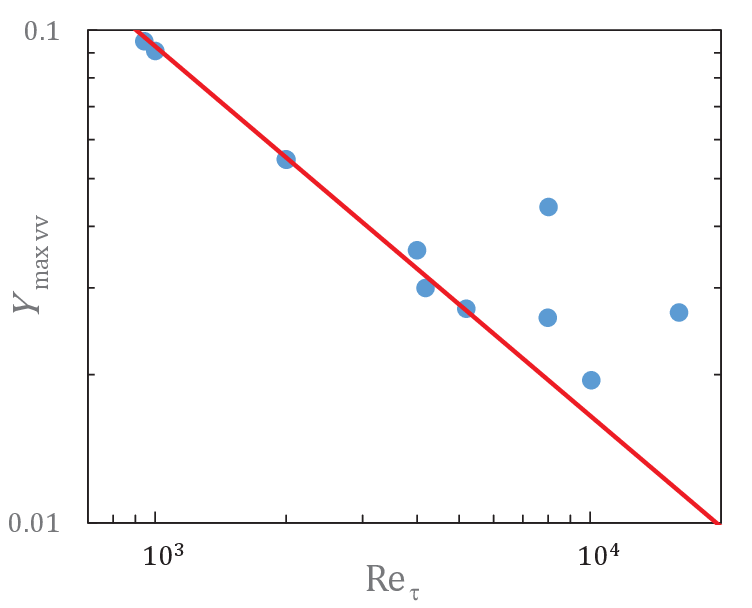}
\caption{\label{Fig12} Location of the maxima (blue $\bullet$) of $\langle vv\rangle$ for the DNS of table \ref{TableDNS}. --- (red), $16.4\,\Reytau^{-3/4}$. }
\end{figure}

Subtracting $\Reytau^{-5/4}$ times equation (\ref{vvin1}) from the DNS leads in figure \ref{Fig11} to a good overall data collapse for $\langle vv\rangle_{in,\,0}$ (again with the exception of profile \#11), which becomes excellent below $y \approxeq 50$. To develop a consistent model for $\langle vv\rangle_{in,\,0}$, its approach to the constant 1.31 at large $y$ needs to be determined. This amounts to finding how the location of the  maximum of $\langle vv\rangle$ scales with $\Reytau$. The Reynolds number scaling of the $\langle vv\rangle$ maxima is shown in figure \ref{Fig12}. Up to $\Reytau$ of 5200, their location is well described by
\begin{equation}
\label{vvmax}
Y_{max\,vv} = 16.5\,\Reytau^{-3/4}
\end{equation}
and the scatter beyond is attributed to increasing uncertainties of the DNS.

The ``local composite expansion'' of $\langle vv\rangle$ in a neighborhood of its maximum around $y \approx 10^2$ is of the form
\begin{equation}
\label{vvmax2}
\langle vv\rangle \approxeq 1.31 - C\,y^{-\alpha} - 500\,\Reytau^{-5/4} - 1.18\,Y^{5/4}\quad ,
\end{equation}
where $C$ and $\alpha$ are to be determined. The maximum of (\ref{vvmax2}) is obtained from its derivative
\begin{equation}
\label{vvmax3}
Y_{max\,vv}^{(5/4 + \alpha)} = \left\{\frac{4\,\alpha\,C}{5\,\times\,1.18}\right\}\,\Reytau^{-\alpha} \quad .
\end{equation}

Combining equations (\ref{vvmax}) and (\ref{vvmax3}) yields $\alpha = 15/4$ and  $C = 4.81\,10^5$, and leads to the Pad\'e approximant
\begin{eqnarray}
\langle vv\rangle_{in, 0} = \{6.2\,10^{-5}\,y^4 + 2.5\,10^{-7}\,y^6 + 6.6\,10^{-9}\,y^8 + 1.2\,10^{-14}\,y^{47/4}\}\quad \nonumber   \\
\times\{1 + 7.7\,10^{-4}\,y^4 + 1.\,10^{-6}\,y^6 +
+ 8.4\,10^{-9}\,y^8 + (1.2/1.31)\,10^{-14}\,y^{47/4}\}^{-1} \label{vv0}
\end{eqnarray}
which has both the appropriate Taylor expansion and the best fit asymptotic expansion. This
completes the full MAE description of $\langle vv\rangle$, summarized by the composite expansion
\begin{equation}
\label{vvcomp}
\langle vv\rangle_{comp} = \langle vv\rangle_{in, 0} + \Reytau^{-5/4}\,\langle vv\rangle_{in,\,-5/4} + W_{vv}(Y)
\end{equation}
given by equations (\ref{vv0}), (\ref{vvin1}) and (\ref{Wvv}).

\section{\label{sec6}The log indicator function for the mean velocity}

Although there exists an extensive literature on turbulent MVPs, this author is not aware of a high fidelity full, meaning from wall to centerline, matched asymptotic description of the channel MVP, which captures its subtle slope variations. To clearly reveal the latter, the logarithmic indicator function
\begin{equation}
\Xi = y\,dU/dy\,=\,Y\,dU/dY   \label{Xilog}
\end{equation}
is reexamined here. To describe the inner part, a sum of exponentials again proves useful, as each simple exponential explicitly reveals the scale of the associated slope change:
\begin{eqnarray}
\Xi_{in}(y) &=& \frac{1}{0.417} + 0.27\,\exp(-0.0007\,y) \label{Xiin1}\\
   &-&0.7\,\exp(-0.009\,y) \label{Xiin2}\\
   &+&9\,\exp(-0.073\,y) \label{Xiin3}\\
   &-&10.97\,\exp(-0.1505\,y\,-9.144\,10^{-3}\,y^2\,-8.612\,10^{-4}\,y^3 \nonumber \\
   &\,& \quad\quad\quad\quad\quad  +1.18\,10^{-4}\,y^4\,-5.\,10^{-6}\,y^5) \label{Xiin4}
\end{eqnarray}
The coefficients of $y$ to $y^4$ in the last exponential (\ref{Xiin4}) are determined such as to produce  the Taylor expansion
\begin{equation}
\label{XiTaylor}
\Xi_{in} = y - 1.295\,10^{-3}\,y^4 + H.O.T.
\end{equation}
about the wall, deduced in \citet[][equ. 7]{Monk24}.

\begin{figure}
\center
\includegraphics[width=0.7\textwidth]{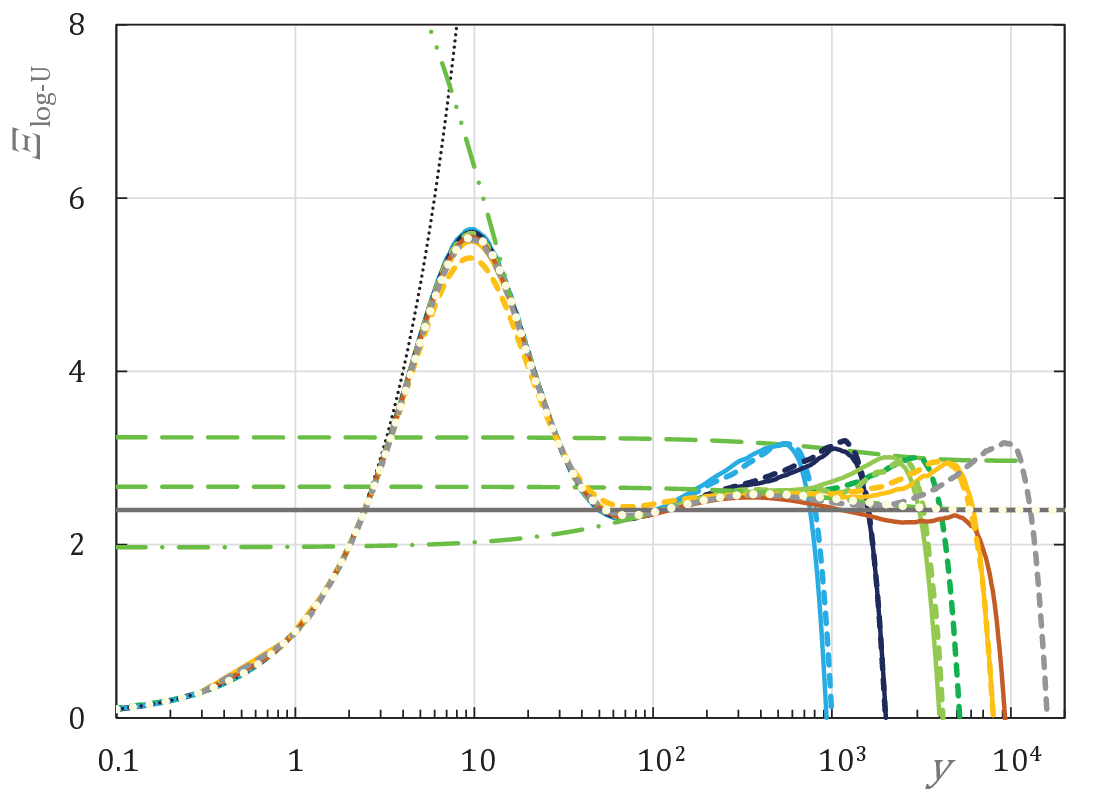}
\caption{\label{Fig13} Logarithmic indicator functions $y\,(d U/d y)$ of the mean velocity for the CFD's of table \ref{TableDNS}. Color and line styles as in the table. --- (grey), $\Xi = 2.4 \,\,(\kappa = 0.417)$; - - - (green), exponential (\ref{Xiin1}) plus its up-shifted image; $-\cdot-\cdot-$ (green), exponential (\ref{Xiin2}); $-\cdot\cdot-\cdot\cdot-$ (green), exponential (\ref{Xiin3}); .... (black), 1-term Taylor expansion (\ref{XiTaylor}); $\bullet\bullet\bullet$ (light yellow), full inner fit (\ref{Xiin1}-\ref{Xiin4}). }
\end{figure}

The fit (\ref{Xiin1}-\ref{Xiin4}) of the inner $\Xi_{in}(y)$ is shown as yellow dots in figure \ref{Fig13} and is seen to provide an exceptionally good model for all but two DNS, up to the point where $\Xi$ develops outer-scaled ``humps'' above the overlap $(1/0.417)$ (the grey horizontal line in figure \ref{Fig13}).
No finite $\Reytau$ corrections to the above  $\Xi_{in}(y)$ are considered here. If there are any, they are of the order of the data uncertainty and cannot be determined reliably.

The outer expansion is modeled as
\begin{eqnarray}
\Xi_{out}(Y) &=& \frac{1}{0.417} \label{Xiout0}\\
&+& 0.05\,\ln\{1+\exp[32(Y-0.13)]\} \label{Xiout1}\\
   &-&C\,\ln\{1+\exp[2000(Y-0.36)^3]\} \quad \mathrm{with} \nonumber\\
   C &=& 1.907\,10^{-3}\, [\Xi_{in}(y=\Reytau)+1.392] \label{Xiout2}
\end{eqnarray}
where the first term (\ref{Xiout0}) is the common part or overlap $(1/0.417)$ of inner and outer expansions and the terms (\ref{Xiout1}) plus (\ref{Xiout2}) constitute the wake function. Its first part (\ref{Xiout1}) represents a quasi-linear departure from the inner expansion, branching off $\Xi_{in}$ at $Y = 0.13$, and the coefficient $C$ of last term (\ref{Xiout2}) is adjusted such that the additive composite expansion
\begin{equation}
\Xi_{comp} = \Xi_{in}(y) + \Xi_{out}(Y) - \frac{1}{0.417} \,, \label{Xicomp}
\end{equation}
is zero on the centerline.

\begin{figure}
\center
\includegraphics[width=0.7\textwidth]{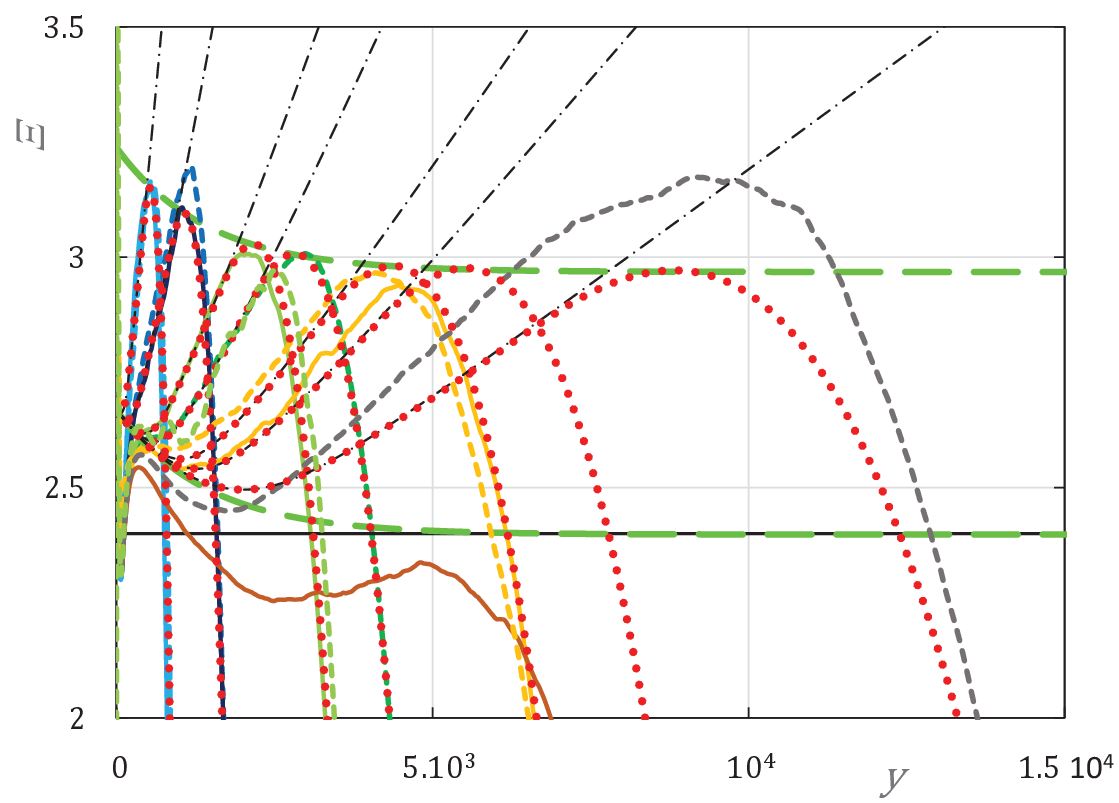}
\caption{\label{Fig14} Same as figure \ref{Fig13} on linear scale. --- (grey), $\Xi = 2.4 \,(\kappa = 0.417)$; - - - (green), exponential (\ref{Xiin1}) plus its up-shifted image; $\bullet\bullet\bullet$ (red), outer quasi-linear parts and decays to the CL; $-\cdot-\cdot$ (thin black), outer quasi-linear parts (\ref{Xiout1}) only. }
\end{figure}

This composite expansion (\ref{Xicomp}) is shown in figure \ref{Fig14} and calls for two comments:
\begin{itemize}
\item The first concerns the quasi-linear part of equation (\ref{Xiout1}), shown in figure \ref{Fig14} as thin black $- \cdot -\cdot -$ lines for the different $\Reytau$. These lines peel off the inner expansion at the fixed outer $Y$ of 0.13 and have no matching inner part. This is because the quasi-linear part in question has been designed to go to zero exponentially for $Y \ll 0.13$ \cite[see e.g.][for a discussion of transcendentally small terms in MAE]{WilcoxP}. Note that the absence of a matching higher order linear term $\propto y/\Reytau$ in the inner expansion, proposed in \citet{Monkewitz_Nagib_2023}, is consistent with the latest conclusions of \citet{Nagib_Marusic_2025}. \newline
    As shown in the above papers and references therein, the slope of the quasi-linear region (\ref{Xiout1}) is strongly flow-dependent. This will be further discussed in the concluding section \ref{sec7}.
\item The second comment concerns the progressively important deviations of the data from the outer fits as $\Reytau$ increases. Up to profile \#7 of table \ref{TableDNS}, the DNS for similar $\Reytau$'s are reasonably well fitted by the composite expansion (\ref{Xicomp}). Beyond $\Reytau$'s of around 5000, however, the computed $\Xi$ become progressively more erratic and the difference between equation (\ref{Xicomp}) and the DNS quickly grows beyond 10\% in the region $y > 10^2$, indicating that the limits of current DNS have been reached for $\Xi$. This is consistent with the demonstration in \citet[][fig. 4]{Monk24}, that the $\Xi$ determined from the computed Reynolds stress $\langle uv\rangle$ and the mean momentum equation differ significantly from figure \ref{Fig14}.\newline
\end{itemize}

\begin{figure}
\center
\includegraphics[width=0.5\textwidth]{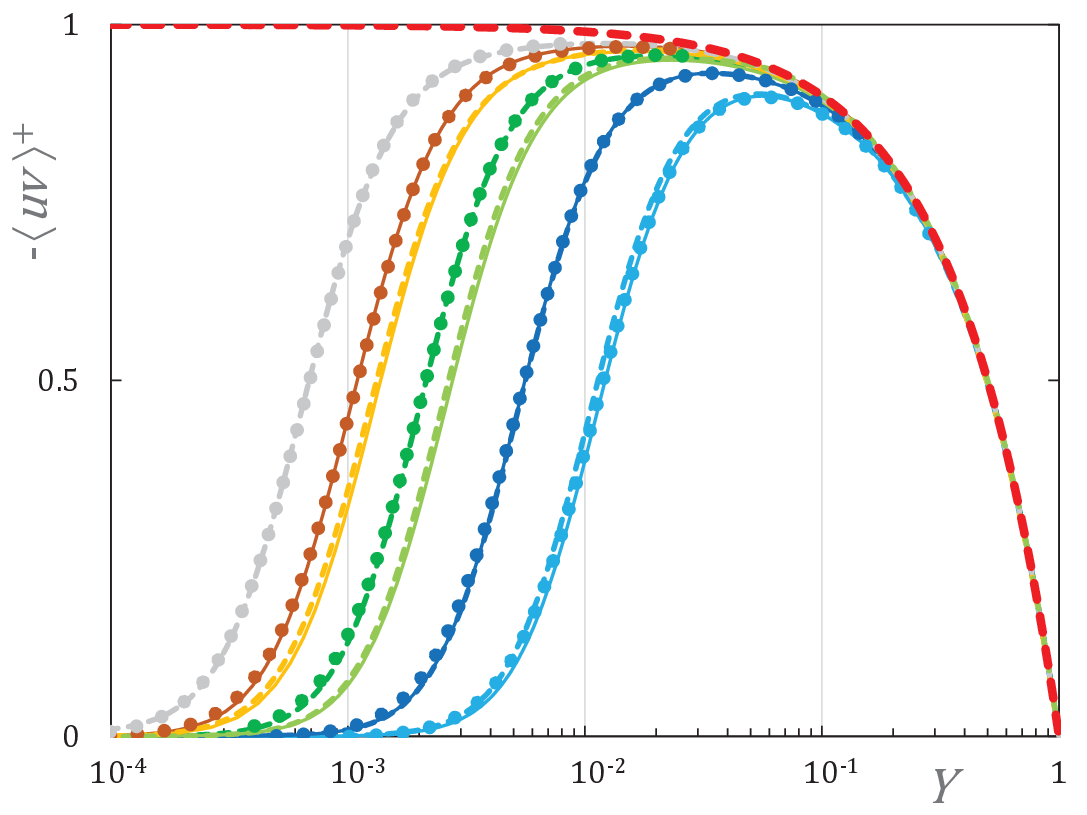}
\caption{\label{Fig15} Negative Reynolds stress $-\langle uv\rangle$ for the DNS of table \ref{TableDNS}. - - - (red), asymptote $(1-Y)$. $\bullet \bullet \bullet$, composite fits of $-\langle uv\rangle$ for the $\Reytau$'s of DNS \#1,4,7,10 and 11, obtained from the fit (\ref{Xicomp}) of $\Xi_{comp}$ and the mean momentum equation.}
\end{figure}

This naturally raises the question of justification for the fit of $\Xi_{out}$ by equations (\ref{Xiout0}-\ref{Xiout2}), especially for large $\Reytau$. A first support comes from inverting the argument of \citet{Monk24} and compute the Reynolds stress from the composite expansion \ref{Xicomp} of $\Xi$ and the mean momentum equation. This is shown in figure \ref{Fig15} and supports the present composite expansion \ref{Xicomp} ``within drawing accuracy'' up to the highest $\Reytau$ available.


\begin{figure}
\center
\includegraphics[width=0.5\textwidth]{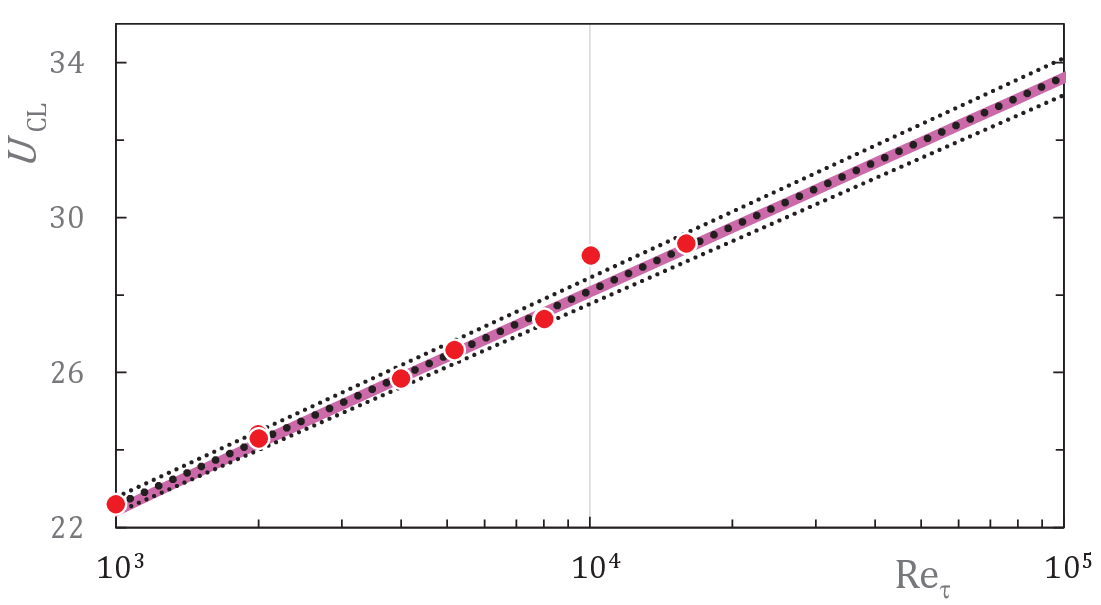}
\caption{\label{Fig16} Centerline velocities for the cases of table \ref{TableDNS}; $\bullet$ (red) DNS; --- (pink), CL velocities from integrating the full fit for $\Xi/y^+$. $\cdot\cdot\cdot$ (black), $6 + 2.4 \ln\Reytau$ with slopes modified by $\pm 2.5\%$. }
\end{figure}

To further validate the composite fit (\ref{Xicomp}), it is divided by $y$ and integrated from wall to centerline in order to obtain the fitted centerline velocity $U_{CL}$, which is shown in figure \ref{Fig16} to be in excellent agreement with the DNS considered here, except again with DNS \#10.

From the above discussion of the channel MVP and from figure \ref{Fig13} in particular, it is evident that, at the $\Reytau$'s available today, it is hazardous to estimate the K\'arm\'an parameter $\kappa$ from the indicator function $\Xi$. This conclusion is of course not new, as it has been reached long ago by \cite{Coles56}, who always advocated using the $\Reytau$ dependence of centerline or free-stream velocity to determine $\kappa$. What is new here, is the determination of the complete $\Xi$ with its oscillations and false ``flats'' that can be and have been mistaken for logarithmic overlaps.

The present analysis also illustrates the power of asymptotic analysis to extract the correct overlap from data at relatively low $\Reytau$, at which the overlap $\Xi_{OL} = 2.4$ is contaminated from both sides by inner and outer terms. To actually start seeing the clean overlap, the inner expansion (\ref{Xiin1}-\ref{Xiin4}) would have to closely approach the overlap (1/0.417), \textit{before} the outer expansion (\ref{Xiout1}-\ref{Xiout2}) starts to significantly deviate from it. This is the case when both $y \gtrapprox 4000$ (see fig. \ref{Fig13}) and $Y \lessapprox 0.08$ (see fig. \ref{Fig14}). Such a scale separation is only attained for $\Reytau > 5.\,10^4$, clearly beyond the reach of DNS in the foreseeable future.

Finally, the spatial oscillations of $\Xi_{in}$ in figure \ref{Fig13} are characterized like those of $\langle uu\rangle_{in,0}$ at the end of section \ref{sec3}. The first three simple exponentials in equations (\ref{Xiin1}-\ref{Xiin3}) yield length scales of $(0.0007)^{-1} \approxeq 1400; \,(0.009)^{-1} \approxeq 110$ and $(0.073)^{-1} \approxeq 14$. The smallest scale of 2.4 is again obtained from the intersection of the leading term $y$ in the Taylor expansion (\ref{XiTaylor}) with the asymptotic value of $\Xi_{in} = 1/0.417$, and is of the order of the viscous sublayer thickness. These observations are compared to the oscillations of $\langle uu\rangle_{in,0}$ in the following concluding section \ref{sec7}.

\section{\label{sec7}Conclusions and some speculations}

The composite asymptotic expansions in sections \ref{sec3}-\ref{sec6} provide the first rigorous MAE descriptions of all first and second order moments in turbulent channel flow obtained from current DNS, strongly supports the ``CS'' scaling of \citet{chen_sreeni2021,chen_sreeni2022,chen_sreeni2023} for $\langle uu\rangle$ and $\langle ww\rangle$, and complements it with the new $1.31 - 1.18\,Y^{-5/4}$ overlap for $\langle vv\rangle$.
Only the Reynolds stress $\langle uv\rangle$ has not been analyzed here, as it is easily obtained from the mean indicator function $\Xi$ and the mean momentum equation, as shown in figure \ref{Fig15}. The inverse, however, i.e. obtaining $\Xi$ from $\langle uv\rangle$ and the momentum equation, is not possible with the currently available DNS, as shown by \citet[][fig. 4]{Monk24}, and one to two additional significant digits will be required beyond $y \approx 50$ to remedy the situation.

Next, a comment on the choice of multiples of $\Reytau^{1/4}$ for the present inner asymptotic expansions of turbulent stresses is required, whereas several authors \citep[for instance][]{Pirozzoli_2024} have advocated various correlations with ``odd'' powers of $\Reytau$. An important point is that these ``odd'' powers have been proposed to fit \textit{total} quantities, while in the present study the $n/4$ powers of $\Reytau$ represent the scaling of \textit{individual terms in asymptotic expansions}. At any rate, the restriction to powers in multiples of $(1/4)$ have not compromised the quality of the present asymptotic expansions. Furthermore, it is difficult to imagine how complex conservation laws, starting with the Reynolds stress transport equations, could be balanced order by order with expansions of the most basic turbulence quantities in terms of ``odd'' powers of $\Reytau$, but the question remains open.

\subsection{\label{sec7a}Comments on normal stresses}

The most unexpected new result of the present study is the $Y$-dependence of the $\langle vv\rangle$ overlap. As it is far from the constant predicted by the attached eddy model, it would be interesting to test it further. Such a test can be devised by taking the incompressible continuity equation in terms of outer-scaled coordinates, multiplying by $v$ and time averaging, which results in
\begin{equation}
\label{cont1}
\langle v\,\frac{\partial u}{\partial X}\rangle + \langle v\,\frac{\partial v}{\partial Y}\rangle + \langle v\,\frac{\partial w}{\partial Z}\rangle = 0
\end{equation}
From this, one obtains in the region of the $\langle vv\rangle$ overlap (\ref{vvOL}) a prediction for the correlation $\langle v \,(\partial u /\partial X)\rangle$ without adjustable parameters
\begin{equation}
\label{cont2}
\frac{1}{2} \, \frac{\dd \langle vv \rangle}{\dd Y} = - \, \langle v\, \frac{\partial u}{\partial X} \rangle = -\, \frac{5}{8} \, 1.18\,Y^{1/4}
\end{equation}

The present first complete inner-outer matched asymptotic expansions should be a good starting point to develop improved models for the near-wall turbulence structures. For this, the spatial extent of the overlaps for the three normal stresses, summarized below, appears particularly relevant :
\begin{eqnarray}
\langle ww\rangle\ Olap:\quad y\approxeq 100\quad \rightarrow \quad Y\approxeq 0.4 \label{OLextww}\\
\langle vv\rangle\ Olap:\quad y\approxeq 150\quad \rightarrow \quad Y\approxeq 0.55 \label{OLextvv}\\
\langle uu\rangle\ Olap:\quad y\approxeq 800\quad \rightarrow \quad Y\approxeq 0.75 \label{OLextuu}
\end{eqnarray}

Finally, the question of universality of the Reynolds stress overlaps must be asked. A priori, no strict universality can be expected, as the overlap represents the link between the near-wall region and a \textit{flow specific} outer region. However, the difference between channel and pipe appears small, as evidenced by comparing figure \ref{figXich} to figure \ref{figpipe} in the appendix.

\subsection{\label{sec7b}Speculation on a possible connection between the oscillations of $\langle uu\rangle$ and of the MVP indicator function}

\begin{figure}
\center
\includegraphics[width=0.4\textwidth]{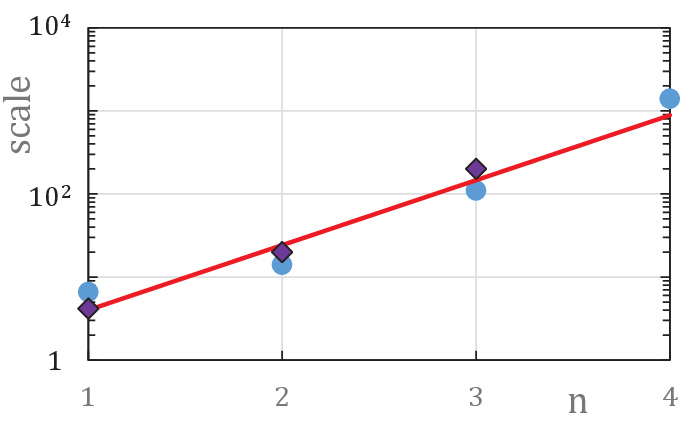}
\caption{\label{Fig17} Consecutive wall-normal scales associated with the oscillations of $\Xi$ and $\langle uu\rangle$ in figures \ref{Fig13} and \ref{Fig4}: $\bullet$ (blue), sequence of oscillation scales for the MVP indicator function $\Xi$ (equ. \ref{Xiin1}-\ref{Xiin4}); $\blacklozenge$ (violet), analogue for $\langle uu\rangle$ (equ. \ref{uu01}-\ref{uu03}). --- (red), speculative fit (\ref{scales}). }
\end{figure}

Also of interest would be a structural model which reproduces the subtle oscillations of $dU/dy$, amplified in figure \ref{Fig13} by the multiplication with $y$, and those of $\langle uu\rangle$ in figure \ref{Fig4}, which may or may not be connected. The spatial scales of these oscillations, obtained from the sequence of exponentials in the fit  (\ref{Xiin1}-\ref{Xiin3}) for $\Xi_{in}$ and from the fit (\ref{uu01}-\ref{uu02}) for $\langle uu\rangle_{in\,,0}$, are shown in figure \ref{Fig17}, together with the speculative correlation
\begin{equation}
\label{scales}
scale = 0.4\,e^{2\,n}
\end{equation}
A connection may exist between these oscillations and the various layers discussed in \cite{WFKM05}, \cite{KFWM07} and \cite{Klewicki_2013-1,Klewicki_2013-2}, but more diagnostics will be required to arrive at a structure based explanation.

\subsection{\label{sec7c}Speculation on the flow dependence of the outer MVP}

Finally, coming back to the MVP, it is recalled that the slope of the ``quasi-linear'' part in equation (\ref{Xiout1}) for the outer indicator function $\Xi_{out}$ (see fig. \ref{Fig14}) is specific to the channel. In pipes this slope is about twice the channel slope, which led \citet{Luchini17} to the conclusion that it is proportional to the mean pressure gradient. The extensive data analysis of \cite{Monkewitz_Nagib_2023} did however not support this exact proportionality, and the physical origin of the quasi-linear region (\ref{Xiout1}) remains to be elucidated.

\begin{figure}
\center
\includegraphics[width=0.6\textwidth]{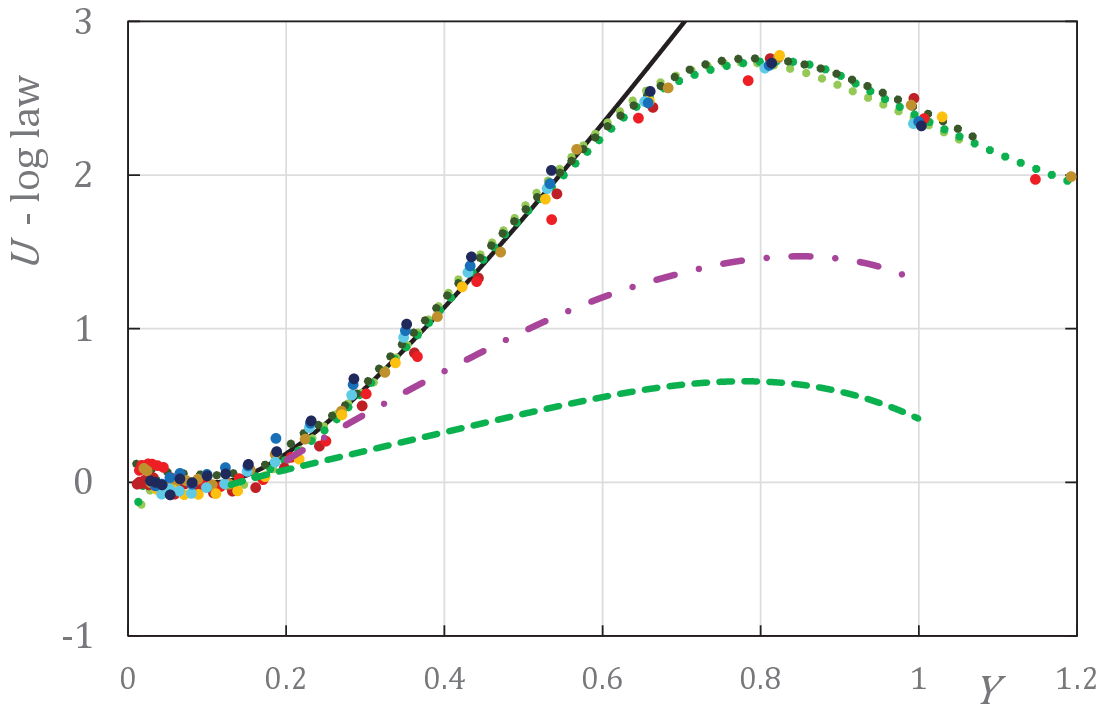}
\caption{\label{Fig18} Mean velocity profiles minus log-laws $(1/\kappa)\,\ln{\Reytau} + B$. $\bullet \bullet \bullet$ (different colors), ZPG TBLs from fig. 7 of \cite{Monkewitz_Nagib_2023} ($\kappa = 0.384,\, B = 4.17$); - - - (green), channel DNS \#7 ($\kappa = 0.417,\, B = 5.65$); $- \cdot - \cdot -$ (violet), pipe DNS of \citet{Pirozzoli_2021} at $\Reytau = 6000$ ($\kappa = 0.433,\, B = 6.65$).}
\end{figure}

The comparison of channel, pipe and ZPG boundary layer MVPs in figure \ref{Fig18} reveals that the ZPG TBL exhibits by far the largest slope of the quasi-linear region of all three MVPs. Hence, a connection between quasi-linear slope and the mean pressure gradient appears tenuous, and a new ``geometric'' explanation is proposed here. Relative to the channel, the span-wise space in the pipe is reduced towards the centerline by a factor $(1-Y)$, which is thought to ``squeeze'' the outer turbulence structures in the $z$ direction, thereby enhancing $\langle uv\rangle$ and $\dd U/\dd y$. In the ZPG TBL, the even stronger enhancement is thought to be caused by the intrusion of high momentum free-stream fluid into the boundary layer \citep[see for instance][]{Chauhan14a}. This basic explanation could possibly be tested in pipes with solid or free slip central cores of varying diameter.

\section{\label{app}Appendix: Brief comparison to the $\langle uu\rangle$ overlaps in pipes}

\begin{figure}
\center
\includegraphics[width=0.48\textwidth]{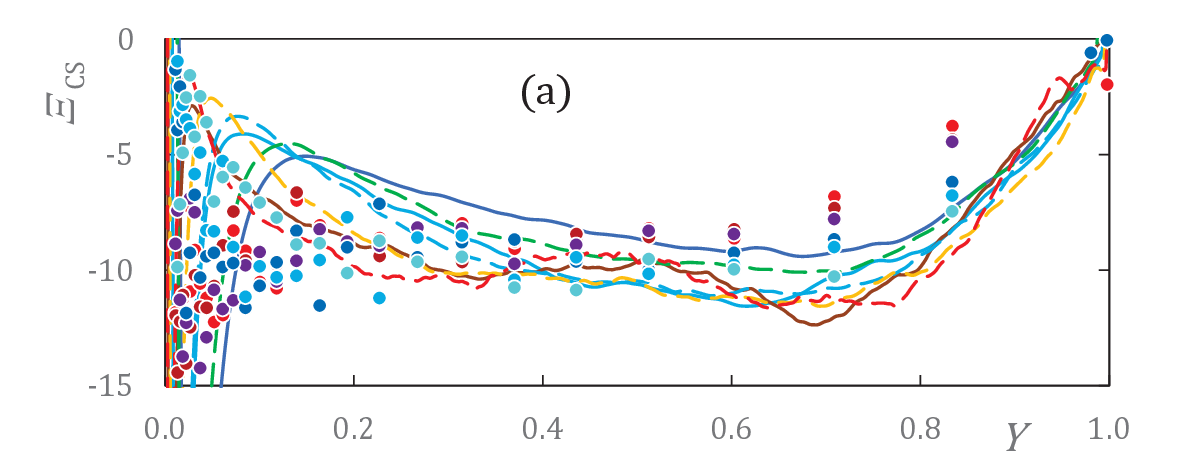}
\includegraphics[width=0.48\textwidth]{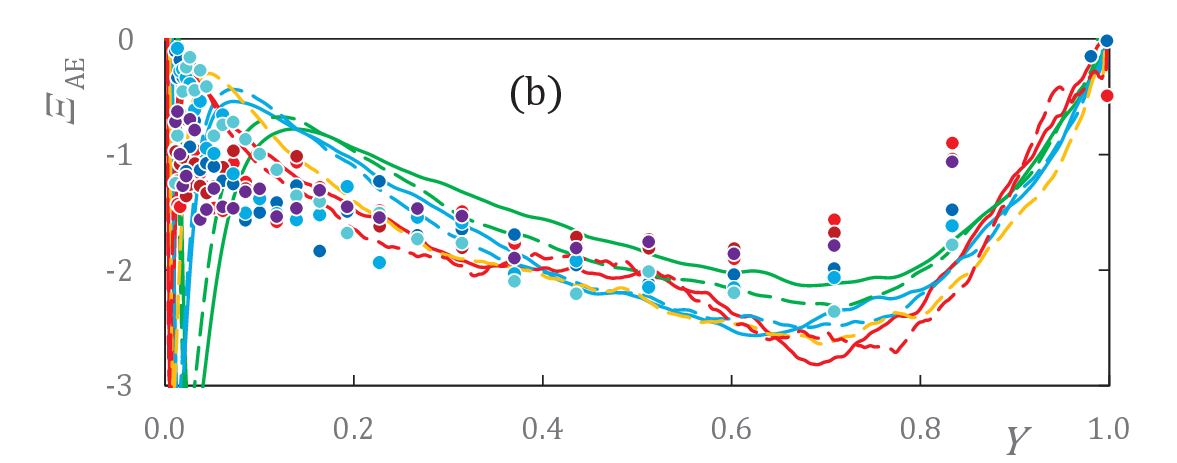}
\caption{\label{figpipe} Pipe flow: CS indicator functions
$\Xi_{\mathrm{CS}}= 4\,Y^{3/4} \dd \langle uu\rangle/\dd Y$  in panel (a) and $\Xi_{\mathrm{AE}}= Y \dd \langle uu\rangle/\dd Y$ in panel (b). $\bullet$ (light, medium, dark blue, violet, dark red, red), smooth Superpipe data of \citet{Hultetal12} for $\Reytau = 5.41, 10.48, 20.25, 37.45, 68.37, 98.19 \times 10^3$; --- (green, blue, red), DNS data of \citet{Yao2023} for $\Reytau = 1.00, 2.00, 5.19 \times 10^3$; - - - (green, blue, orange, red), DNS data of \citet{Pirozzoli_2021} for $\Reytau = 1.14, 1.98, 3.03, 6.02 \times 10^3$.}
\end{figure}

For pipe flow, $\Xi_{\mathrm{CS}}= 4\,Y^{3/4} \dd \langle uu\rangle/\dd Y$ and $\Xi_{\mathrm{AE}}= Y \dd \langle uu\rangle/\dd Y$ have been evaluated for the smooth Superpipe data of \citet{Hultetal12} and for selected DNS profiles of \citet{Pirozzoli_2021} and \citet{Yao2023}, all shown in figure \ref{figpipe}. As seen in panel (a), the data closely follow the indicator value of -10 for the CS channel overlap law (\ref{uuOLBD}) up to $Y \approx 0.4-0.5$. Beyond this $Y$, the Superpipe data return to zero considerably faster than the DNS, which is surprising as probe corrections generally diminish towards the centerline. At the same time, some pipe DNS are seen to yield somewhat erratic $\Xi$'s at the higher $\Reytau$, which has motivated the focus on channel DNS in this paper. Nevertheless, the CS scaling is seen to produce a region of reasonably constant $\Xi_{CS}$ in figure \ref{figpipe}(a), while no such region appears in figure \ref{figpipe}b.

The author is grateful to Yoshinobu Yamamoto for sharing his unpublished Channel DNS. \newline


\bibliographystyle{jfm}
\bibliography{Turbulence}

\end{document}